\documentclass{aa}
\def\micron{$\mu\mathrm{m}$\,}
\def\kms{$\mathrm{km\,s}^{-1}$\,}

\def\HH{{\mathrm H_2}}
\def\HF{\mathrm{HF}}
\def\CH{\mathrm{CH}}

\usepackage{natbib,graphicx}
\usepackage{txfonts}
\begin{document} 

   \title{Unveiling the chemistry of interstellar CH\thanks{The reduced spectra (Fig.~\ref{Fig2}) and Table 2 are available in
electronic form (in fits and ASCII format, respectively) at the CDS via
anonymous ftp to cdsarc.u-strasbg.fr (130.79.128.5) or via
http://cdsweb.u-strasbg.fr/cgi-bin/qcat?J/A+A/}}
   \subtitle{Spectroscopy of the 2~THz $N=2\leftarrow 1$ ground state line}

   \author{H. Wiesemeyer
          \inst{1}
          \and
          R. G{\"u}sten
          \inst{1}
          \and
          K.M. Menten
          \inst{1}
          \and
          C.A. Dur{\'a}n
          \inst{1}
          \and
          T. Csengeri
          \inst{1}
          \and
          A.M. Jacob
          \inst{1}
          \and
          R. Simon
          \inst{2}
          \and
          J. Stutzki
          \inst{2}
          \and
          F. Wyrowski 
          \inst{1}
          }

   \institute{Max-Planck-Institut f{\"u}r Radioastronomie, Auf dem H{\"u}gel 69,
              53121 Bonn, Germany\\
              \email{hwiese@mpifr.de}
           \and
              I. Physikalisches Institut, Universit{\"a}t zu K{\"o}ln,
              Z{\"u}lpicher Str. 77, 50937 K{\"o}ln, Germany
             }

   \date{Received August 22, 2017; accepted December 12, 2017}

 
  \abstract
   {}
   {The methylidyne radical CH is commonly used as a proxy for molecular hydrogen in the cold,
    neutral phase of the interstellar medium. The optical spectroscopy of CH
    is limited by interstellar extinction, whereas far-infrared observations provide an integral
    view through the Galaxy. While the HF ground state absorption, another H$_2$ proxy
    in diffuse gas, frequently suffers from saturation, CH remains transparent both in spiral-arm
    crossings and high-mass star forming regions, turning this light hydride into
    a universal surrogate for H$_2$. However, in slow shocks and in regions
    dissipating turbulence its abundance is expected to be enhanced by an endothermic production
    path, and the idea of a ``canonical'' CH abundance needs to be addressed. 
   }
   {The $N=2\leftarrow 1$ ground state transition of CH at $\lambda 149$~\micron has become
    accessible to high-resolution spectroscopy thanks to GREAT, the {\it German Receiver for Astronomy at
Terahertz Frequencies}
    aboard the {\it Stratospheric Observatory for Infrared Astronomy}, SOFIA. Its
    unsaturated absorption and the absence of emission from the star forming regions makes it an ideal candidate
    for the determination of column densities with a minimum of assumptions. Here we present
    an analysis of four sightlines towards distant Galactic star forming regions, whose 
    hot cores emit a strong far-infrared dust continuum serving as background signal. Moreover, if
    combined with the sub-millimeter line of CH at $\lambda 560$~\micron,
    environments forming massive stars can be analyzed. For this we present a case
    study on the ``proto-Trapezium'' cluster W3~IRS5.
    }
   {While we confirm the global correlation between the column densities
    of HF and those of CH, both in arm and interarm regions, clear signposts
    of an over-abundance of CH are observed towards lower densities. However,
    a significant correlation between the column densities of CH and HF
    remains.  A characterization of the hot cores in the W3~IRS5 proto-cluster and its envelope
    demonstrates that the sub-millimeter/far-infrared lines of CH reliably trace not only
    diffuse but also dense, molecular gas.
   }
   {In diffuse gas, at lower densities a quiescent ion-neutral chemistry alone cannot account
    for the observed abundance of CH. Unlike the production of HF, for $\mathrm{CH}^+$ and CH, vortices
    forming in turbulent, diffuse gas may be the setting for an enhanced production path. However,
    CH remains a valuable tracer for molecular gas in environments reaching from diffuse clouds to
    sites of high-mass star formation.
   }

   \keywords{ISM: abundances -- clouds -- lines and bands -- molecules -- structure}

   \maketitle
%

\section{Introduction}

The methylidyne radical, CH, was the first molecule observed in space 
\citep{Dunham1937} and identified by \citet{Swings1937}.
Most CH originates in the diffuse gas of the cold, neutral matter (CNM) via the slow
radiative association of C$^+$ with H$_2$ forming CH$_2^+$, which converts to
CH via dissociative recombination CH$_2^+$(e$^-$,H)CH
\citep[e.g.,][]{Gerin2016}. Owing to the former reaction, one expects
the CH abundance to scale with that of H$_2$, making CH a proxy for molecular
hydrogen. Other reaction channels are the dissociative recombination of
CH$_3^+$, and the photodissociation of CH$_2$. The usefulness of CH as a
surrogate for H$_2$ is confirmed by \citet{Sheffer2008}, who observed the
CH $A-X (0-0)$ band at 430~nm. They derive CH column
densities that range from $\sim$10$^{11}$ to $\sim$10$^{13}\,\mathrm{cm^{-2}}$.
The reference column densities of H$_2$ were derived from UV spectra of the
Lyman $B-X$ bands. However, UV and optical studies are restricted to
distances of up to $\sim$7~kpc, with the vast majority being much smaller \citep[e.g.,][]{Sheffer2008}, owing to Galactic extinction in the UV, and to the
requirement of bright background stars in the optical.
\citet{Pan2005} found that there are two populations of CH bearing clouds, one 
being associated with CN, the other one with CH$^+$. The abundance of CH$^+$ and
subsequently CH may indeed be enhanced by the molecular hydrogen abstraction reaction
C$^+$(H$_2$,H)CH$^+$. This reaction is endothermic by $0.398 \pm 0.003$~eV or about
4620~K \citep[][while earlier measurements yielded  $\sim 4300$~K,
\citealp{Gerlich1987}]{Hierl1997}. The supply activating this reaction is thought to be provided
by the dissipation of turbulence and by ion-neutral drift \citep{Godard2009} or by slow,
non-dissociative shocks permeating the CNM \citep[][further references
therein]{Gerin2016}. Indeed, \citet{Godard2014} demonstrated that at 
$n_{\rm H} = 50$~cm$^{-3}$, 80\% to 90\% of the CH column density and all of the
CH$^+$ arise from bursts of dissipation, while the remaining fraction is
associated with gas in the ambient medium or relaxing towards it. Only at gas densities
in excess of typically 100~cm$^{-3}$ does the abundance of CH become nearly constant
\citep{Sheffer2008,Levrier2012}. As a matter of fact, at $n_{\rm H} = 300$~cm$^{-3}$,
about
70\% of the CH originates in ambient conditions \citep{Godard2009}.

These results notwithstanding, CH is commonly used as a proxy for H$_2$. It shares this virtue
with HF, which was first detected in the ISM by \citet{Neufeld1997}. Because HF is the only neutral
light hydride that exothermically forms in a triatomic reaction with H$_2$ \citep{Neufeld2005,Zhu2002},
it is considered the more pristine H$_2$ tracer. However, unlike for CH, its ground-state absorption often saturates,
not only towards the dense molecular gas associated with star-forming regions. The same holds for
the ground-state absorption of OH at 2.5~THz which can be considered a secondary H$_2$ tracer
\citep[e.g.,][]{Wiesemeyer2016}. This leaves CH as an important proxy at high column densities.
In the more diffuse gas, below $A_{\rm V} = 1$, all three species are more reliable
H$_2$ tracers than CO, which suffers from insufficient self-shielding \citep[][further references therein]{Gerin2016}.

These findings leave their fingerprint in the measured abundances
and their mutual correlations. Thanks to the operation of spaceborne
and airborne far-infrared instrumentation with adequate spectral resolution
(HIFI and GREAT\footnote{GREAT is a development by the MPI für Radioastronomie and the
KOSMA/ Universität zu Köln, in cooperation with the MPI für Sonnensystemforschung
and the DLR Institut für Planetenforschung.}
aboard the Herschel Space Observatory\footnote{Herschel is an ESA space observatory with science instruments provided by European-led Principal Investigator consortia and with important participation from NASA.} and SOFIA, respectively),
such studies can now be conducted Galaxy-wide, unlike UV and
optical observations. The spectroscopy of ground-state
lines of the aforementioned light hydrides, observed in absorption along
sightlines towards the hot, far-infrared bright dust associated with sites of
high-mass star formation (for many of which distances have been determined from maser parallaxes, e.g.,
\citealp{Reid2014}), provides a clear-cut setting to determine the column density
profiles of spiral-arm crossings with a minimum of assumptions:
As an indication, only $\sim$100~ppm of the total population in diffuse clouds
resides in excited states (for details we refer to Appendix \ref{app:nlte}, in particular
Table~\ref{table:A1}). Historically, besides the aforementioned UV/optical studies,
CH column densities were determined using the radio transitions between the
hyperfine-split ground-state levels \citep[e.g.,][]{Liszt2002}. Although
the involved level populations are affected by non-LTE level occupations,
the error in the derived column densities remains moderate if the transition is optically
thin and inverted, and if the uncertain excitation temperature is expected to obey the condition
$|T_\mathrm{ex}| \gg T_\mathrm{bg}$ (where $T_{\rm bg}$ is the temperature of the continuum
background against which the radio spectrum is observed, and assuming that beam filling
factors are sufficiently well known to ensure that this condition holds). The non-LTE
effect is briefly discussed in Appendix \ref{app:nlte}.

\begin{figure}[ht!]
\centering
\resizebox{\columnwidth}{!}{\includegraphics[]{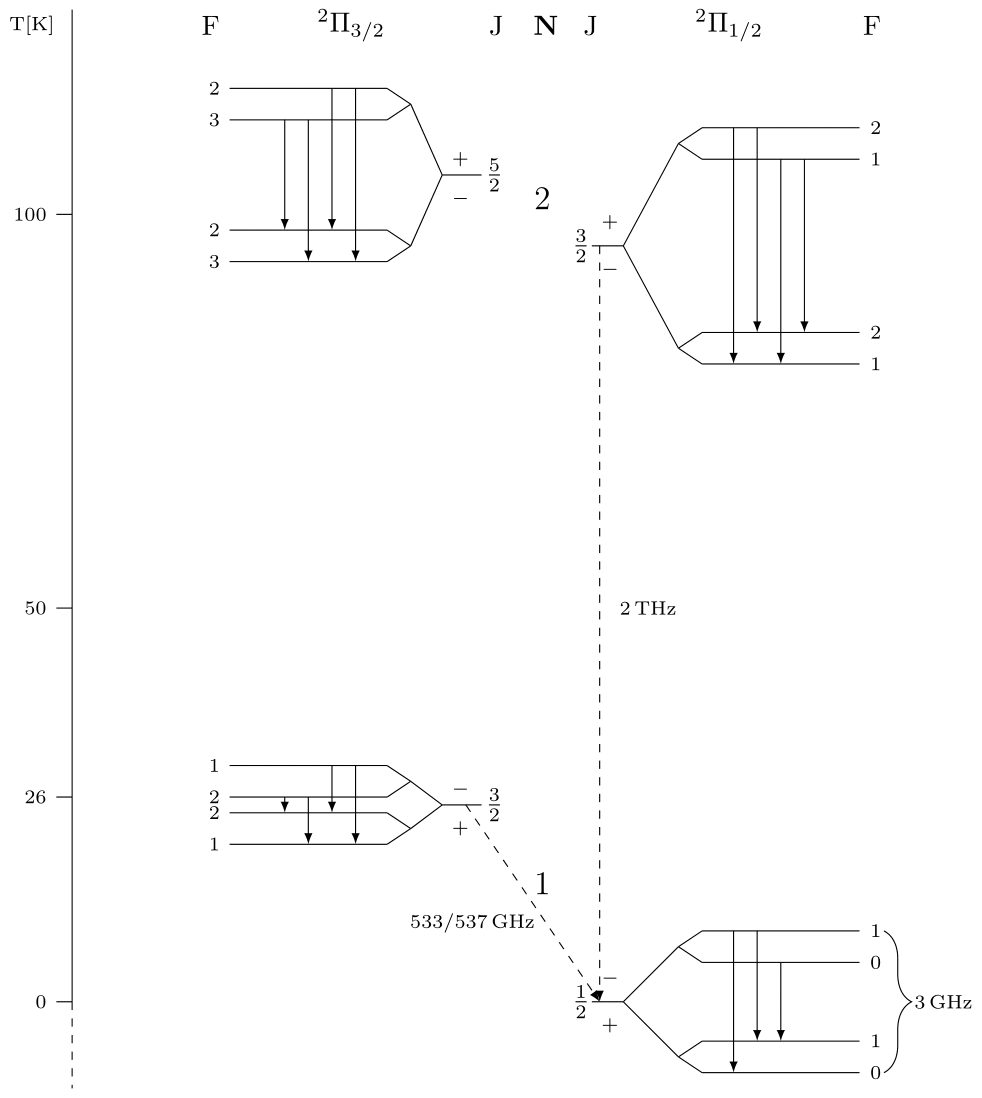}}
\caption{The lowest rotational energy levels of the CH radical. The sub-millimeter/far-infrared transitions
discussed here, at 533/537 GHz (i.e., $\lambda 560$~\micron) and 2007/2010 GHz ($\lambda 149$~\micron) are
indicated by the dashed arrows. Hyperfine splitting is not to scale.\label{fig:levels}}
\end{figure}

The installation of the upGREAT array and its extended tuning range
\citep{Risacher2016} made the transition of CH molecules from the $N=1, J=1/2$ ground state into
the $N=2, J=3/2$ state accessible to observations (the energy level diagram
of CH is shown in Fig.~\ref{fig:levels}). This transition comprises two triplets of hyperfine structure
lines with wavelengths near $149~\mu$m, two of which were first detected with very low spectral
resolution with the Kuiper Airborne Observatory \citep{Stacey1987}. With adequate spectral resolution to
distinguish the features originating in different spiral arms, this line lends itself particularly well
to absorption studies of diffuse and translucent clouds because it remains in absorption even in the
background source. The rotational $\lambda 560$~\micron transition between the
$N=1,\,J=3/2\rightarrow 1/2$ shares its lower level with the ground-state transition studied here,
and was detected with HIFI at high spectral resolution \citep{Gerin2010}. Because the $\lambda 560$~\micron
line of CH is easier to excite than its counterpart at $\lambda 149$~\micron (cf. Fig.~\ref{fig:levels}),
in the environment of the background source it appears in emission, which, by virtue of its hyperfine splitting
(cf. Table~\ref{table:1}), blends with the sight-line features from spiral arms and interarm gas, rendering
the analysis more difficult. In a similar context, the detection of the
$A-X(0,0)$ band of CH near 4304~{\AA} in the environment of Herschel 36, the
O-star primarily illuminating the hourglass nebula M8, reveals that its lower
level, the excited $^2\Pi, J=3/2$ fine structure state 25.6~K above ground, is
significantly populated \citep{Oka2013}.

The aim of this work is therefore to establish the $\lambda 149$~\micron line as a
tool allowing us to obtain CH (and ultimately $\HH$) column densities from first
principles, with a minimum of a priori knowledge
required. We will compare CH column density
profiles with those of HF \citep[][further references therein]{Sonnentrucker2015}
and OH \citep{Wiesemeyer2016}, so as to assess the usefulness of CH as a tracer for molecular
gas in diffuse clouds. We extend our analysis to the envelopes around regions
forming high-mass stars, where, despite the high densities, the CH absorptions at
$\lambda 149$~\micron and $\lambda 560$~\micron do not saturate. When analyzed together with
the emission in the $\lambda 560$~\micron line, CH remains a suitable $\HH$ tracer in such
environments.

\begin{table*}
\caption{Hyperfine splitting of the $N=2\leftarrow 1, J=3/2\leftarrow 1/2$
transitions.}
\label{table:1}
\centering
\begin{tabular}{cccccc}
\hline\hline
HFC & Frequency & $A_{\rm E}$          & $E_{\rm L}$ & $E_{\rm U}$ & $n_{\rm crit}$ \\   
    &   [GHz]   & $[\mathrm{s}^{-1}]$  &     [K]     &     [K]     & $[\mathrm{cm}^{-3}]$ \\   
\hline
$F = 1^- \leftarrow 1^+$ & 2006.74892 & 0.01117 & 0.00072 & 96.31011 & \\
$F = 1^- \leftarrow 0^+$ & 2006.76263 & 0.02234 & 0.00000 & 96.31005 & $\sim 3\times 10^{10}$ \\
$F = 2^- \leftarrow 1^+$ & 2006.79912 & 0.03350 & 0.00072 & 96.31252 & \\
\hline
$F = 1^- \leftarrow 1^+$ & 2010.73859 & 0.01129 & 0.16071 & 96.66158 & \\
$F = 1^- \leftarrow 0^+$ & 2010.81046 & 0.02257 & 0.15726 & 96.66157 & $\sim 2\times 10^{10}$ \\
$F = 2^- \leftarrow 1^+$ & 2010.81192 & 0.03386 & 0.16071 & 96.66510 & \\
\hline                              
\end{tabular}
\tablefoot{Data are from \citet{Pickett1998}. Col. (1): Hyperfine component. Col. (3): Einstein coefficient. Cols. (4), (5): lower and upper state
energy. Col. (6): Critical density, $A_{\rm E}/\gamma$, for 100~K gas temperature. The collisional rate
cofficients $\gamma$ do not account for hyperfine splitting and are scaled by reduced mass
\citep[from He to para-$\HH$ as collision partner,][]{Marinakis2015}.}
\end{table*}
\section{Observations and data analysis}
\label{sec:observations}
The observations were performed on several SOFIA flights in the observatory's
cycle 4 (F298 and F301 on 2016 May 18 and 24, respectively, F346 on 2016 
November 8, and F369 on 2017 February 3). The aircraft altitudes were 41.7 to 42.6 
kft. The configuration of GREAT was 
single-pixel channel L1 and upGreat's arrays LFA-H and -V (seven pixels each).
The frequency of the $N = 2 \leftarrow1, J = 3/2 \leftarrow 1/2$
transition of CH at 2006.762~GHz (Table~\ref{table:1}) was tuned to
the upper sideband, so as to avoid a prominent telluric ozone feature appearing 
in the image band of a lower sideband tuning. Typical single-sideband receiver
temperatures were 1650 to 1750~K, while the signal-band opacity varied from 0.2 to 0.3.
The spectra were taken in chop-nod mode, with a chop amplitude of $80''$ and
a frequency of 2.5~Hz. The CH absorption spectrum appears in the central
pixel of the LFA-H array, while its remaining six pixels, covering a hexagonal arrangement
with $33''$ spacing \citep[][this corresponds to 2.3 times the beam width (full width at half maximum, FWHM)
at the observing frequency]{Risacher2016} were used to better define the atmospheric
total power emission off the source. After down-conversion to an intermediate-frequency
bandpass of 4~GHz, the raw data were analyzed with XFFTS spectrometers providing 283~kHz
resolution, covering the bandpass with $2^{14}$ channels \citep{Klein2012}. The spectra were
calibrated with the {\sc kalibrate} program \citep{Guan2012}, which is part of the {\sc kosma}
package also used to execute the observations. The underlying forward
and main-beam efficiencies are 0.97 and 0.68, respectively. Further processing with the
{\sc class} software \footnote{Development led by IRAM, {\it http://www.iram.fr/IRAMFR/GILDAS/}.}
included spectral smoothing to 0.36~\kms -wide velocity bins, which is a
good compromise between sensitivity and resolving power for narrow absorption features, and a
correction for spectral baselines by removing polynomials of up to third order.
The continuum level, determined by means of a dedicated double-sideband calibration, was added back
to the spectra in order to ensure a correct line-to-continuum ratio.

In contrast to techniques applied for optical/UV spectroscopy, the
analysis of rotational ground state lines arising in diffuse clouds
needs no excitation or extinction corrections. At their typical temperatures of
15 to 100~K, one can safely assume that the CH radical is almost entirely
in its $^2\Pi_{1/2}$ ground state (cf. Appendix~\ref{app:nlte}), and spontaneous de-excitation
can be neglected. The opacity for a velocity component $i$ and a hyperfine component
$j$ therefore reads
\begin{equation}
\tau_\mathrm{ij, \upsilon} =
\sqrt{\frac{\ln{2}}{\pi}} \frac{A_{\mathrm E,j}c^3}{4\pi\Delta \upsilon_{\rm i}
\nu_{\rm j}^3}
\frac{g_{\rm u,j}}{g_{\rm l,j}} N w_{\rm j}
\exp{
     \left(-4\ln{2}\left(\frac{\upsilon-\upsilon_{0,ij}}
                              {\Delta \upsilon_i}\right)^2\right)
    }\,,
\label{eq:1}
\end{equation}
where $A_{\mathrm E,j}$ is the Einstein coefficient and
$\Delta \upsilon_{\rm i}$ the half-power width of the Gaussian profile of the
velocity component centered at $\upsilon_{0,ij}$. The degeneracies of the 
upper and lower level of hyperfine components $j$ involved in the
transition are denoted $g_{\rm u,j}$ and
$g_{\rm l,j}$, respectively. $N$ is the total column density of CH in the ground
state, and the factor
$w_{\rm j} = g_{\rm l,j}\exp{(-E_{\rm l,j}/(kT))}/Q(T)$ describes the fractional 
population in the hyperfine-split level $j$, at an energy $E_{\rm l,j}$ above ground
and for an excitation temperature $T$, with partition function $Q(T)$.

In the following analysis, we fit ensembles of Gaussian velocity components to the
opacity spectra obtained from the observations, that is, each component follows the ansatz given by
Eq.~\ref{eq:1}. The numerical implementation of this method is described in
\citet{Wiesemeyer2016}. In view of the saturated absorption in portions of the absorption spectra
of OH and HF, we emphasize that such an approach implies two features: Entirely saturated velocity
components do not arise in the analysis, because they can evidently not constrain the minimization
of the merit function. On the other hand, components of which only a minor portion is affected by
saturated absorption still constrain the column density across the full velocity interval (but only
for the cloud entity that they represent).
The propagation of the baseline noise and of the uncertainty in the line-to-continuum
ratio into the determination of column densities is discussed in Appendix~\ref{app:err}.
A note on the treatment of asymmetric error distributions is given in Appendix~\ref{app:asymerr}.
\section{Results}
\subsection{Main characteristics of the observed sightlines}
\begin{figure*}
   \centering
   \resizebox{0.9\hsize}{!}
      {\includegraphics{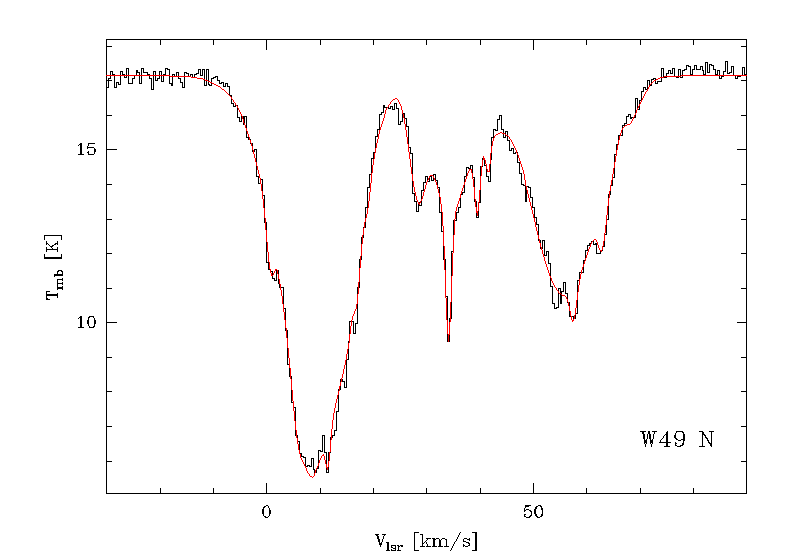}\includegraphics{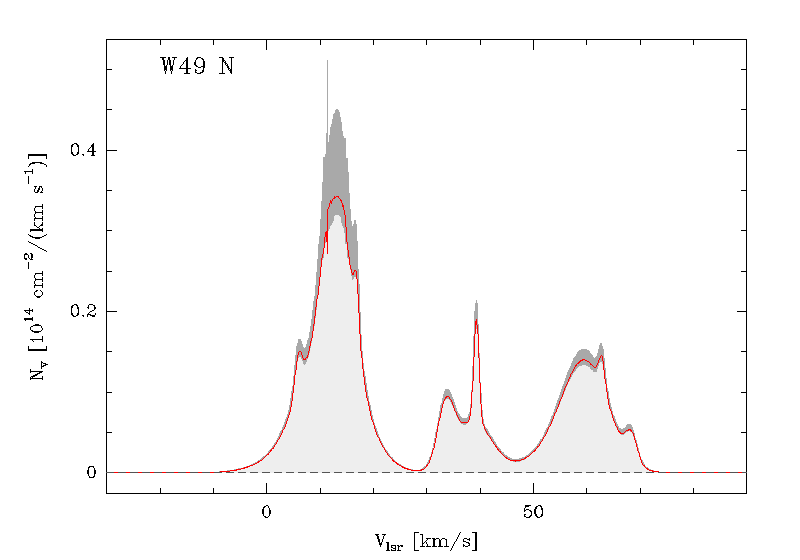}}
   \resizebox{0.9\hsize}{!}
      {\includegraphics{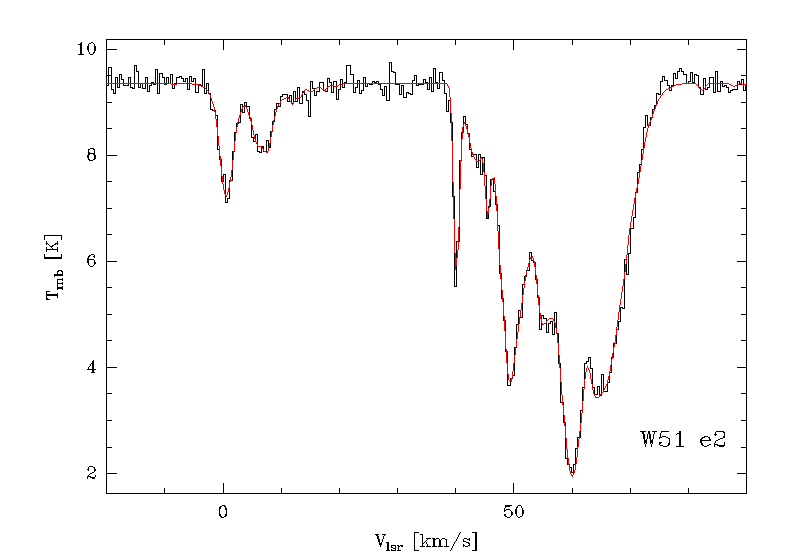}\includegraphics{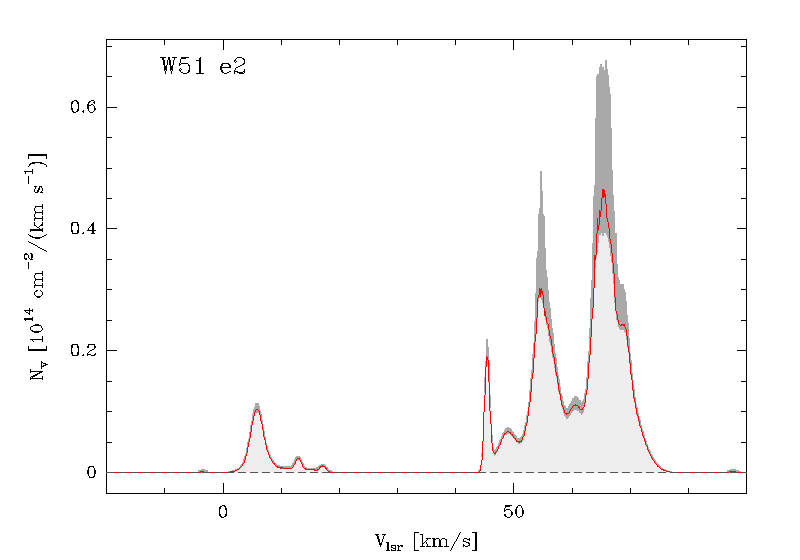}}
   \resizebox{0.9\hsize}{!}
      {\includegraphics{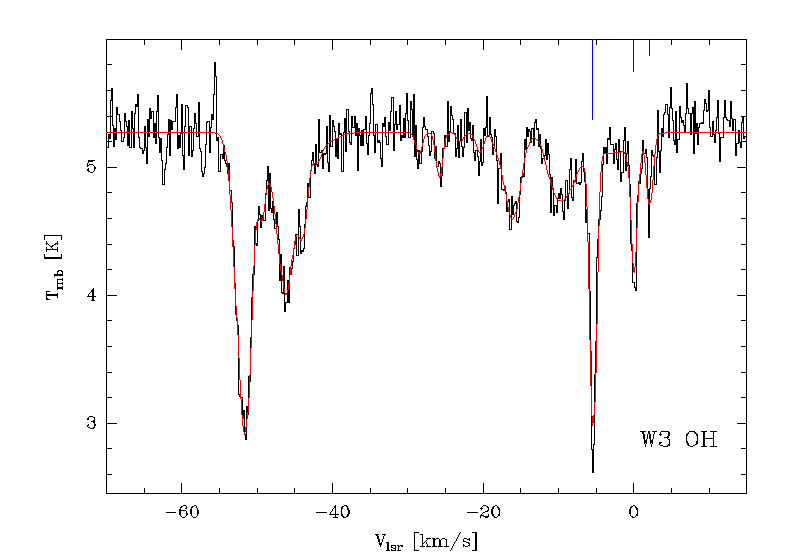}\includegraphics{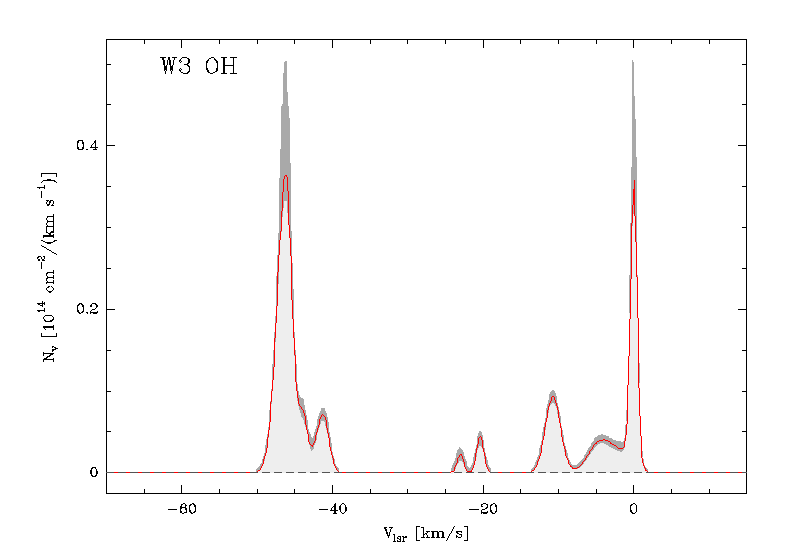}}
   \resizebox{0.9\hsize}{!}
      {\includegraphics{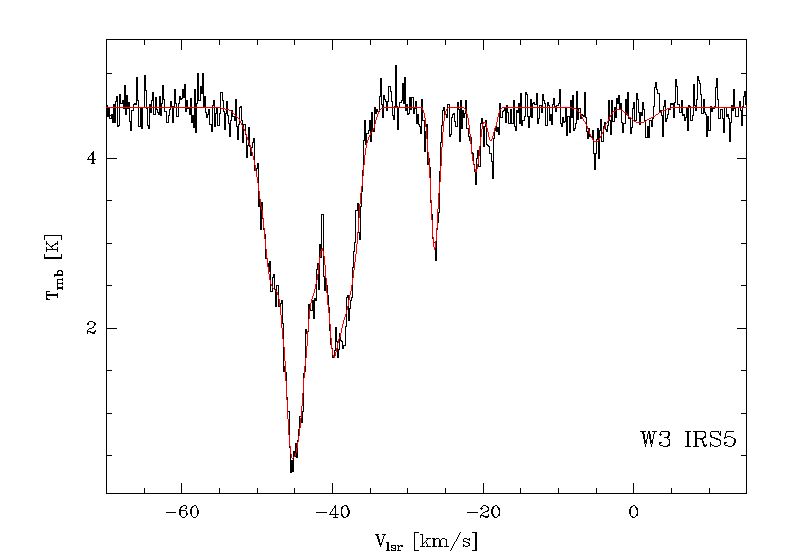}\includegraphics{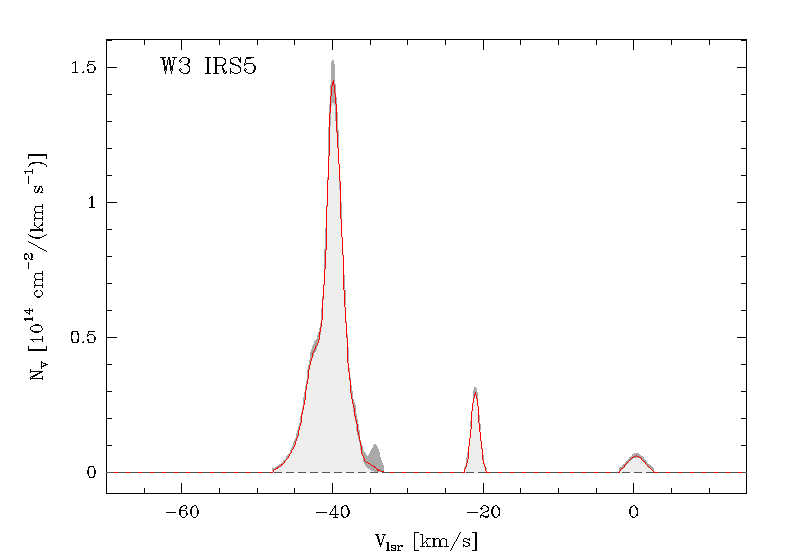}}
   \caption{CH $\lambda 149\,\,\mu$m spectra (left, the reference frequency
            for the velocity scale given by the second HFC) and deduced
            column density profiles (right) for (from top to bottom): 
            W49\,N, W51\,e2, W3(OH), W3\,IRS5. The blue markers in the W3(OH)
            spectrum show the hyperfine splitting and weights). The median column densities are
            indicated by the red lines. The dark-gray-shadowed
            profiles indicate the $(-\sigma_-, +\sigma_+)$ limits,
that is, 84.13\% confidence, assuming a 5\% error in the continuum
            level. \label{Fig2}}
   \end{figure*}

The spectra of the $\lambda 149\,\,\mu$m transition of CH are shown in
Fig.~\ref{Fig2}, along with the deduced velocity distributions of the column density
(typically $10^{13}$~cm$^{-2}$). Before we proceed to their quantitative analysis, we
briefly summarize the main characteristics of the observed sightlines.

The mini-starburst template W49\,N, forming a young massive cluster \citep{Galvan-Madrid2013},
is located in the Perseus spiral arm, at 11.1~kpc distance \citep{Zhang2013}. The sightline grazes
the Sagittarius spiral arm, which displays absorption at velocities from 30 to 70~\kms, in
two groups corresponding to the near and far side. The velocity interval from $-10$ to
$+30$~\kms represents the Perseus spiral arm. Being close to zero velocity,
these absorption components may also contain gas from the local arm, a conjecture corroborated by
the sightline to W51~e1/e2 discussed below. Thanks to the $\sim$80~\kms wide
velocity range and its length, this sightline provides a good basis for abundance studies 
with solid statistics (e.g., \citealp{Monje2011} for the HF $1-0$ line, and \citealp{Gerin2010}
for the $\lambda 560\,\,\mu$m CH doublet), while it avoids the complexity of the central molecular
zone of the Galaxy. We therefore used this sightline to confirm that the calibration of the
line-to-continuum ratios in the upGREAT and HIFI spectra is consistent (Appendix~\ref{app:chch}).

The second target, W51~e1/e2 (which is part of the W51~A region), is a similar protocluster,
though less massive than W49\,N \citep[e.g.,][]{Ginsburg2017}. At a similar Galactic longitude,
its distance of 5.4~kpc \citep{Sato2010} locates this complex in the near-side crossing of the Sagittarius
spiral arm, where the CH column is distributed across a $\sim$30~\kms -wide velocity range, separated
from the gas in the 0 to 20~\kms interval by a 25~\kms -wide gap void of CH. The CH column across
these low velocities must arise from gas located closer to, or located in, the solar neighborhood. It cannot
arise from the Perseus arm as in the case of W49\,N, given the shorter distance to W51~e1/e2.
The presence of CH in this velocity component is not surprising, since it was known to be traced by other molecules
or molecular ions like CH$^+$ \citep{Falgarone2010b}, OH$^+$ and OH \citep{Wiesemeyer2016},and HF and H$_2$O
\citep{Sonnentrucker2015}. The component consists of partially molecular, diffuse gas, since it is also
present in the atomic tracers HI \citep{Winkel2017}, CII \citep{Gerin2012,Gerin2015} and OI \citep{Guesten2017}.

The W3 complex is located in the second quadrant of the Galaxy in the Perseus arm. It consists of a
classical H{\sc ii} region/radio continuum source associated with a prominent far infrared source. Its
most prominent, high-mass star forming sub-regions are W3(OH) and W3~Main
\citep[for an overview see e.g.,][]{Megeath2008}; in projection they are $\sim$10~pc apart.
The archetypical ultracompact H{\sc ii} region W3(OH) \citep[e.g.,][]{Qin2016} is located at a distance of
2~kpc \citep{Xu2006,Hachisuka2006}. W3~Main contains several luminous infared sources, out of which
W3\,IRS\,5, harboring a massive protocluster \citep{Wang2013}, is the IR-brightest. Far-infrared continuum
emission from both sources has been imaged with FORCAST and SOFIA \citep{Salgado2012,Hirsch2012}.
The velocity distribution of CH towards W3(OH) shows two major components, of which one is attributed to the
target itself and its envelope, and the other one to the local spiral arm. Towards W3\,IRS\,5, this latter
component is much less prominent. The sightlines to both targets contain a component close to $-20$~\kms, 
which is attributed to the near side of the Perseus arm \citep[e.g.,][]{Vallee2008}.
\subsection{Quantitative analysis}
\begin{figure*}
   \centering
   \resizebox{\hsize}{!}{\includegraphics{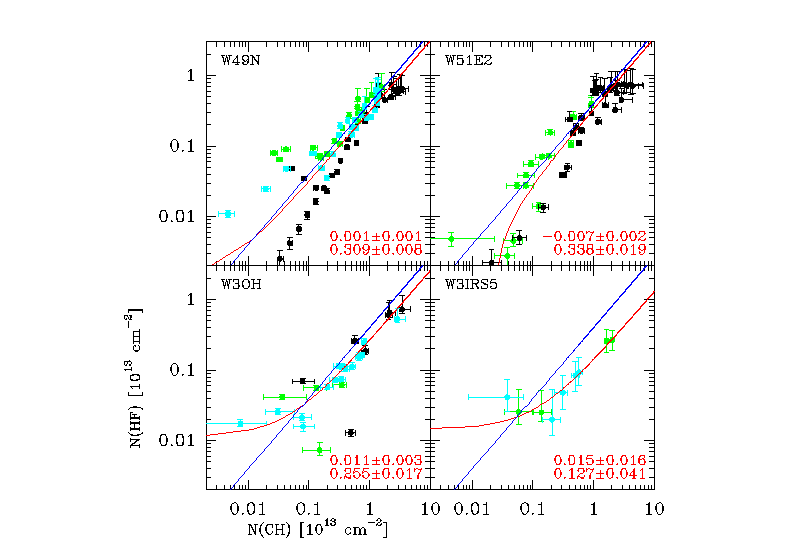}}
   \caption{Correlation between HF and CH column densities
            towards W49\,N, W51\,e2, W3\,IRS5 and W3(OH) (clockwise from
            top left). Column densities are for 1~\kms wide velocity
            intervals (Fig.~\ref{Fig2}). Ordinate offsets and slopes
            from the weighted regression (parameters $a$ and $b$ in Eq.~\ref{eq:weightedRegression})
            are given in the lower-right corners.
            The red lines show the corresponding fits, the blue lines a ratio of
            HF/CH~$=0.4$ \citep{Sheffer2008,Sonnentrucker2015}.
            Double-logarithmic scales illustrate the bimodal distribution 
            at low column densities towards W49\,N and W51\,e2. Black markers
            indicate the absorption around, or close to, the location of the continuum source (W49~N:
            $\upsilon_\mathrm{lsr} < 27$~\kms; W51~e2:
            $\upsilon_\mathrm{lsr} > 30$~\kms; W3(OH), W3\,IRS5:
            $\upsilon_\mathrm{lsr} < -30$~\kms). Green markers indicate:
            For W49~N, $27\,\mathrm{km\,s^{-1}} < \upsilon_\mathrm{lsr} < 47\,\mathrm{km\,s^{-1}}$,
            for W51~e2, $\upsilon_\mathrm{lsr} < 30$~\kms, and for W3(OH), W3\,IRS5 
            $-30\,\mathrm{km\,s^{-1}} < \upsilon_\mathrm{lsr} < -15\,\mathrm{km\,s^{-1}}$.
            Remaining velocities are shown in cyan.
            \label{Fig3}}
   \end{figure*}

In the following analysis we use the HF ground-state absorption as reference
for the available $\HH$ reservoir, which we will correlate with the column densities of CH.
As already stated in the introduction, the exothermic direct association
of F with $\HH$ \citep{Neufeld2005,Zhu2002} lends HF its virtue as primary $\HH$ tracer. 
Here we use archival HF data from the Herschel guaranteed time key program PRISMAS
\footnote{http://astro.ens.fr/?prismas} (P.I. Maryvonne Gerin). The analysis (Fig.~\ref{Fig3}) yields $N(\mathrm{HF})/N(\mathrm{CH})$
ratios of $0.30\pm 0.01$, $0.34 \pm 0.02$, $0.26\pm 0.02$ and $0.13 \pm 0.04$ towards W49\,N, W51\,e2, W3(OH)
and W3\,IRS\,5, respectively. The corresponding correlation coefficients are 0.87, 0.85, 0.94 and 0.98. 

For the evaluation of false-alarm probabilities various statistical tests exist (i.e., for the likelihood that a chance coincidence
mimics a correlation between the column densities of HF and CH). If we interpreted Pearson's $p$ value as false-alarm
probability against our null hypothesis (the column densities of CH and HF being uncorrelated),
our results would be significant at the 1\% level. However, the usefulness of $p$
value testing has been questioned by a controversy which became known as the ``$p$ value fallacy''
\citep{Sellke2001}. We therefore provide an independent test: The non-zero column densities of CH are randomly
permuted among the available velocity intervals (of 1~\kms width each), using a modified Fisher-Yates algorithm.
We then count the number of false alarms (i.e., correlations detected with a $p$ threshold of 0.05) out of a
total of 1000 Monte-Carlo tests. This yields false-alarm probabilities of 4.0, 4.2, 6.0 and 7.9\%, respectively.
If these probabilities were $p$ values, the correlations towards W3(OH) and W3\,IRS5 would merely be considered
``quasi-significant''. However, given that our test is more critical, it seems fair to say that all four sightlines
support the conjecture that HF and CH form out of a common H$_2$ reservoir. Assuming that the average
physical conditions on these sightlines are comparable, one can also stack the data. This yields a highly
significant $p$ value ($\ll 0.01$\%), while the Monte Carlo analysis detects a false alarm in 4.4\% of the sample,
which renders the correlation significant. We performed a dedicated regression analysis, accounting for the
individual errors in $N(\mathrm{HF})$ and $N(\mathrm{CH})$ using the merit function
\begin{equation}
\chi^2 = \sum_{i=1}^{n} \frac{(N_\mathrm{HF,i}-a-bN_\mathrm{CH,i})^2}{\sigma_\mathrm{HF,i}^2+b^2\sigma_\mathrm{CH,i}^2}
\label{eq:weightedRegression}
,\end{equation}
\citep[][their equation (15.3.2)]{Press1992} where $n$ is the number of spectral channels, and 
$\sigma_\mathrm{CH,i}$ and $\sigma_\mathrm{HF,i}$ are the root mean square (r.m.s.) errors in the individual spectral channels, as given
by the distributions resulting from the Bayesian estimate. The analysis reveals that the scatter in the data is not due 
to sensitivity limitations or uncertainties in the line-to-continuum ratio (see Appendices \ref{app:err} and
\ref{app:asymerr}), but rather due to the chemical diversity on the sightlines. The consequences of this finding will be
discussed in the following Section.

The abundance of HF was determined by \citet{Indriolo2013} using the 2.5~$\mu$m ro-vibrational transition.
\citet{Sonnentrucker2015} compared the distribution of HF with that of H$_2$O on twelve sightlines (further references
can be found in these studies); their models predict HF abundances (with respect to $\HH$) of $X(\mathrm{HF}) \sim 0.9 \times 10^{-8}$
in the low-density regime (at $A_{\rm V} = 0.9$) to $X(\mathrm{HF}) = 3.3 \times 10^{-8}$ at $A_{\rm V} \sim 4$ (for 
further references to determinations of $X(\mathrm{HF})$ see \citealp{Wiesemeyer2016}). With these values, our mean HF/CH ratio
of 0.3 translates 
to $X(\mathrm{CH}) = 0.3$ to $1.1\times 10^{-7}$. For comparison, the ``canonical'' CH-H$_2$ relationship,
derived from optical and UV spectroscopy, is $N(\mathrm{CH})/N(\mathrm{H_2}) = 3.5_{-1.4}^{+2.1}\times 10^{-8}$ \citep{Sheffer2008}.
In both studies the uncertainties reflect the chemical diversity along the sightlines, not the measurement errors.
The analysis of \citet{Indriolo2013} and \citet{Sonnentrucker2015} suggests a mean abundance ratio of
$N(\HF)/N(\CH) = 0.4$. Some of our data, at $N(\mathrm{CH})\gtrsim 0.1\times 10^{13}\,\mathrm{cm^{-2}}$, agree with
this relation (blue lines in Fig.~\ref{Fig3}). However, on the sightlines to W49~N and W51~e2, the correlation widens
for $N(\mathrm{HF})\lesssim 0.1\times 10^{13}\,\mathrm{cm^{-2}}$. We suggest that this is due to a
distinctly different chemistry, leading to an over-abundance of CH. We emphasize that the assumption of an overly low
excitation temperature for the velocity intervals coinciding with the star-forming regions in the background is not the reason for this bimodal
distribution: First, using a higher, more realistic
excitation towards the hot cores (cf. Section \ref{sec:sfr}) would simply move the data points upwards along 
the bisecting line without considerably changing the apparent bimodality. In the spectrum
of W49~N, this bimodality is most striking for the data points with
$N(\mathrm{CH})\sim 0.1\times 10^{13}\,\mathrm{cm^{-2}}$, derived from sightline velocities below
$\upsilon_{\rm lsr} < -1.0$~\kms. The negative velocities towards W49~N are known to partially originate
from expanding gas driven by an energetic outflow, evidence supported not only by the wide line-wings seen in HCN
\citep{Liu2015}, but also by the OH $^2\Pi_{1/2},\,J=3/2 \rightarrow 1/2$ emission \citep{Menten2018}
and whose blue wing coincides with less prominent p-H$_2$O emission \citep{Sonnentrucker2015}.
Notwithstanding, it is plausible that we also see unrelated (e.g., local) diffuse foreground gas at the
negative velocities: First, if the absorption there was contaminated by the blue-shifted line wing
related to the outflow-driven expanding gas, we would expect to also see its red-shifted counterpart
(traced by the prominent emission from the excited OH line). This is obviously not the case; at
$+28$~\kms, between the Perseus and Sagittarius spiral arms, the CH column density drops to almost zero.
Second, towards W51\,e2 we also find three data points that exhibit a statistically significant
over-abundance and that do not belong to the gas envelope in which W51~e2 is embedded. As already
mentioned, they are rather located in the diffuse gas of the Sagittarius spiral arm where the
assumption of a complete ground-state population is realistic, or at least inconsequential.
\section{Discussion}
\subsection{Chemistry of CH in diffuse gas}\label{sec:diffuse}
\begin{figure}
   \centering
   \resizebox{\columnwidth}{!}{\includegraphics{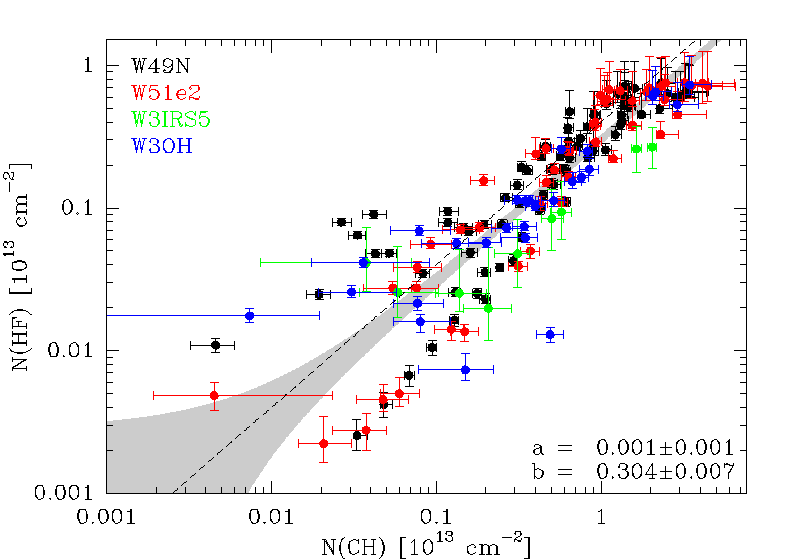}}
   \caption{As in Fig.~\ref{Fig3}, but for all data. A legend for the colors is given in the
            upper-left corner. The gray-shaded area covers all data points within
            the 3$\sigma$ limits of the weighted linear regression.
            \label{Fig4}}
   \end{figure}

The most compelling explanation for the overabundance of CH with respect to its canonical
value is the endothermic hydrogen abstraction reaction forming CH$^+$ from C$^+$, followed
by the exothermic hydrogen abstraction reactions leading to CH$_2^+$ and CH$_3^+$, from where
dissociative recombinations lead to the formation of CH and CH$_2$, respectively. The mystery
around the ubiquity of CH$^+$, identified in the CNM by means of absorption spectroscopy
in optical lines \citep[$^1\Pi \rightarrow ^1\Sigma$ transitions, ][]{Douglas1941} and in the
far-infrared \citep[$^1\Sigma,\, J=1\rightarrow 0$ transition, ][]{Naylor2010,Falgarone2010b} was long ago discussed
in the literature \citep[e.g.,][]{Black1975,Black1977}. CH$^+$ is also found
to be highly abundant in the Orion bar \citep{Naylor2010,Parikka2017}, a prototypical PDR,
and in dense, strongly UV-illuminated gas in regions forming high-mass stars
\citep{Naylor2010,Falgarone2010a}.
It is worth noting that vibrationally excited H$_2$ makes the reaction C$^+$(H$_2$,H)CH$^+$
exothermic \citep{Hierl1997,Zanchet2013}, and that the vibrational excitation of H$_2$ may be chemically pumped thanks to its 
formation on carbonaceous dust grains \citep{Gough1996}. However, while this reaction channel
plays a significant role in photo-dissociation regions \citep{Agundez2010,Faure2017}, it cannot
account for the large CH$^+$ abundance in diffuse and translucent clouds \citep{Agundez2010}.
For these media, one mechanism suggested to overcome the endothermicity of 4620~K is shock
chemistry \citep{PineauDesForets1986,Draine1978}. Another mechanism proposed for the high abundance
of CH$^+$ throughout the CNM and compact HII regions \citep{Naylor2010} is the dissipation of
turbulence \citep{Godard2009}. In this scenario, the synthesis of CH$^+$ and, subsequently, CH is
more efficient for lower gas densities, where more and larger vortices form in the turbulent, diffuse gas,
with an increased lifetime. On the other hand, the majority of the HF column density
builds up in ambient diffuse gas not exposed to dissipating turbulent cells or slow
shocks \citep{Godard2014}. A closer look at the correlation between the abundances of HF and CH
(Fig.~\ref{Fig3}) reveals a substantial fraction of material at
$N(\mathrm{CH}) \lesssim 10^{13}$~cm$^{-2}$ that would not be expected solely from the
corresponding HF abundance. This conclusion is corroborated by the analysis of the stacked data
shown in Fig.~\ref{Fig4}. On the other hand, the resulting regression coefficient $a$ does not significantly
deviate from zero. Based on this data it therefore seems fair to conclude that CH and HF always coexist,
but with abundance ratios varying within the limits that become apparent in Figs.~\ref{Fig3} and \ref{Fig4}.
This variation is also a caveat against an overinterpretation of the stacking analysis.

Correlating the CH overabundance, defined as $0.4 N(\mathrm{CH})/N(\mathrm{HF})$, with $N_{\rm H_2}$, or,
for lack of it, with $N(\mathrm{HF})$ as substitute, indeed yields a fall-off of the CH overabundance with increasing column
density - as expected if one associates larger column densities with larger volume densities,
rather than with sightline crowding. The thus defined overabundance is shown
in Fig.~\ref{Fig6}, against $1.65\times 10^{11}\,\mathrm{cm}^{-2}$ and 
$3.3\times 10^{11}\,\mathrm{cm}^{-2}$ -wide HF column density intervals (for W49\,N and W51\,e2,
respectively). We limit the analysis to column densities
below the threshold where the bimodality becomes apparent. 
The width of the intervals is chosen so as to obtain a statistically
meaningful result, as confirmed by the error bars derived from the variance
of the overabundance within each interval. We selected only
data points with at least 3~$\sigma$ determinations of both CH and HF column densities, which excludes
the portions of the HF spectrum affected by saturated absorption. The
overabundances observed towards W49\,N and W51\,e2 level off from values of $3-4$ to the canonical
value of $\sim$1 when the H$_2$ column density increases by a factor of $\sim$5. Interestingly,
\citet{Godard2009}, exploring the parameter space for the dissipation of turbulence, find similar 
ratios for a model with $A_{\rm V}=0.4$ (after correcting for the varying molecular hydrogen content),
while at $A_{\rm V}=0.1$ the CH abundance varies only mildly. On the other hand, a single sightline
contains diffuse cloud entities of markedly different characteristics, ranging from vortices embedded
in a turbulent medium, with varying molecular content, to the precursors of molecular clouds. It seems
fair to say that this statement holds throughout the whole Galaxy, not only in its central molecular
zone, where these contrasting characteristics are much more pronounced
\citep[e.g.,][]{Ginsburg2016}. Moreover, these manifestations of diffuse, cold gas may blend with each
other even within a given spiral arm. Therefore, more spiral-arm crossings need to be investigated,
including so far unexplored sightlines, because only solid statistical evidence can corroborate the
theoretical work in a meaningful way.

Previous studies remain inconclusive regarding an over-abundance of CH. The dual-slope relationship
between the column densities of CO and CH, with a break close to 
$N(\mathrm{CH}) \sim 10^{13}\,\mathrm{cm}^{-2}$, is attributed to increasing self-shielding of CO above
the break point \citep{Sheffer2008}, and has nothing to do with a variation of the abundance of CH.
\begin{table*}
\caption{Synopsis of velocity-integrated column densities of CH, HF, and OH on the sightline to W49\,N.
Error estimates are based on the normalized $\chi^2$ of the fits shown in Fig.~\ref{Fig2}, and on a 5\% uncertainty
in the continuum levels.
\label{table:2}}
\centering
\begin{tabular}{l c c c c c c c}
\hline\hline
               & $\upsilon$            & $N(\mathrm{CH})$     & $N(\mathrm{HF})$ & $N(\mathrm{OH})$ & $f_{\rm H_2}$ & $N(\mathrm{HF})/N(\mathrm{CH})$ & $N(\mathrm{OH})/N(\mathrm{CH})$  \\
               & [\kms]  & \multicolumn{3}{c}{$[10^{13}\,\,\mathrm{cm}^{-2}]$} & $^{(d)}$ & $^{(e)}$ & $^{(e)}$ \\
\hline
 & & & & \\
Local gas      & $(-3,+5)$   & $ 2.97_{-0.02}^{+0.03}$  & $0.689_{-0.007}^{+0.009}$ & $21.7_{-0.6}^{+1.9}$ & $0.32_{-0.14}^{+0.14}$ & $0.232_{-0.003}^{+0.004}$ & $7.31_{-0.24}^{+0.65}$ \\
 & & & & & & & \\
Perseus$^{(a)}$& $(+5,+20)$  & $ 33.2_{-0.2}^{+1.1}$    & $7.98_{-0.08}^{+0.35}$    & $61.3_{-1.9}^{+5.7}$ & $0.78_{-0.57}^{+0.16}$ & $0.240_{-0.003}^{+0.013}$ & $1.85_{-0.07}^{+0.19}$ \\
 & & & & & & & \\
Interarm       & $(+20,+30)$ & $ 1.67_{-0.02}^{+0.02}$  & $0.580_{-0.004}^{+0.004}$ & $5.99_{-0.14}^{+0.21}$ & $0.31_{-0.22}^{+0.16}$ & $0.347_{-0.005}^{+0.005}$ & $3.59_{-0.09}^{+0.13}$ \\
 & & & & & & & \\
Sgr$^{(b)}$    & $(+30,+45)$ & $ 9.91_{-0.06}^{+0.10}$  & $4.62_{-0.04}^{+0.18}$    & $37.9_{-0.8}^{+1.7}$ & $0.76_{-0.30}^{+0.20}$ & $0.466_{-0.005}^{+0.019}$ & $3.82_{-0.08}^{+0.18}$ \\
 & & & & & & & \\
Sgr$^{(c)}$    & $(+45,+70)$ & $18.35_{-0.09}^{+0.14}$  & $6.47_{-0.05}^{+0.20}$    & $38.1_{-0.4}^{+1.5}$ & $0.54_{-0.14}^{+0.42}$ & $0.353_{-0.003}^{+0.011}$ & $2.08_{-0.03}^{+0.08}$\\
 & & & & & & & \\
\hline
\end{tabular}
\tablefoot{(a) Analysis and error estimates inaccurate (velocity interval contains hot-core environment of
unknown excitation, HF absorption is partially saturated). (b) Near- and far-side crossing of Sagittarius spiral arm.
(c) Far-side crossing. (d) Molecular hydrogen fractions
$f_{\mathrm{H}_2} = 2N(\mathrm{H_2})/(N(\mathrm{HI})+2N(\mathrm{H}_2))$ are derived from HF (with
$N(\mathrm{HF})/N(\mathrm{H}_2) = 1.4\times 10^{-8}$) and from HI $\lambda21$~cm data,
\citep[][further references therein]{Winkel2017}. (e) Bayesian error estimates (accounting for the correlation between
the column densities of the reported ratios).}
\end{table*}
   \begin{figure}[h!]
   \centering
   \resizebox{\columnwidth}{!}{\includegraphics{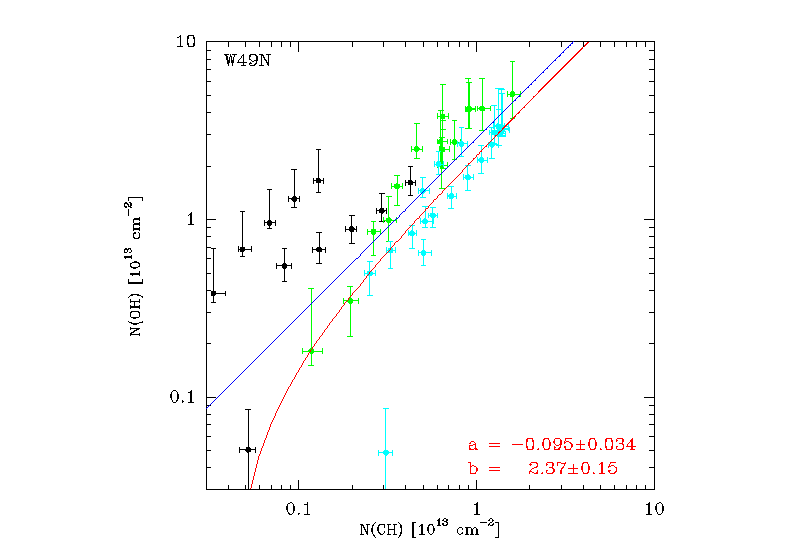}}
   \caption{Correlation between OH and CH column densities
            towards W49\,N. The blue line is for a $N(\mathrm{OH})/N(\mathrm{H_2})$ ratio
            of $10^{-7}$. The red line shows the weighted linear regression, the resulting
            OH/CH abundance ratio is given in the lower-right corner. Velocity bins for
            the column densities and color-codes are as in Fig.~\ref{Fig3}.
            \label{Fig5}}
   \end{figure}
The crossings of Galactic spiral arms are traced as clearly in CH as they are in HF
(Table~\ref{table:2}) and OH \citep{Wiesemeyer2016}. For instance, the contrast between the far-side crossing
of the Sagittarius spiral arm and the adjacent interarm region in front of the Perseus arm is $\sim$10 in CH
and HF, but only $\sim$2 in molecular hydrogen fraction. This can be easily explained by the fact that
W49\,N is located at a galactocentric distance of 7.6~kpc, where atomic gas is more abundant relative to
molecular gas than on sightlines towards the inner Galaxy. A similar contrast has been found in OH, which
can be considered a secondary molecular hydrogen tracer \citep{Gerin2016}; \citet{Mookerjea2016}
indeed find no correlation of the $N(\mathrm{OH})/N(\mathrm{CH})$ abundance ratio with the molecular
hydrogen fraction, suggesting that both molecules trace H$_2$ equally well, whatever
proportion of the total hydrogen budget applies, and that the observed scatter in this abundance ratio
is rather due to other factors, for example, a varying cosmic ray ionization rate. With our data we can confirm
this conjecture: Figure~\ref{Fig5} shows the correlation between the column densities of OH and CH towards
W49\,N (excluding velocity components where the entire OH absorption profile saturates). The correlation has a coefficient
($r$-value) of 0.67 (falling below the $r$-value for the CH-HF correlations), a $p$-value of 1\% and is
significant (the Monte-Carlo analysis detects a false alarm probability of 2.7\%).
The weighted linear regression predicts an OH/CH abundance ratio
of $2.37 \pm 0.15$, which, for the \citet{Sheffer2008} relationship
$N(\mathrm{CH})/N(\mathrm{H_2})=3.5\times 10^{-8}$, converts to
$N(\mathrm{OH})/N(\mathrm{H_2})=8.3\times 10^{-8}$, reasonably close to both predicted and
measured OH abundances
\citep[][respectively]{Albertsson2014,Wiesemeyer2016}. Except for one outlier, the scatter in the OH/HF
abundance ratio is not larger than the scatter in the CH/HF ratio. As a consequence, the over-abundance of CH at low
column densities should also appear when, instead of HF, OH is used as H$_2$ tracer. This is indeed
suggested by the distribution shown in Fig.~\ref{Fig7} (totally absorbed velocity
components are not considered in this analysis, as in the case of HF). However, owing to the larger errors, the evidence
is less compelling. As a reminder, we recall that even within a narrow velocity interval all correlated
quantities are averages of diffuse and translucent cloud entities along the sightline, exposed to varying
conditions. As a matter of fact, the abundances of CH, OH, and H$_2$O are particularly sensitive to the cosmic ray ionization rate
\citep{Godard2014}. One may also speculate that the correlation between CH and OH is less pronounced than that between
CH and HF because at low densities ($n_{\rm H} \la 30$~cm$^{-3}$) OH has a production channel
in the relaxation stage following a burst of turbulent dissipation, as shown by \citet{Godard2014}: While the OH abundance falls below that of CH immediately after the
burst, the OH/CH ratio exceeds unity during the subsequent relaxation stage, on a
timescale that depends on the dissipated energy per unit length.
It is interesting to note that \citet{Liszt2002} determine (across a limited $N(\mathrm{H}_2)$
range) a ratio of $N(\mathrm{OH})/N(\mathrm{CH}) = 3.0 \pm 0.9$, in good agreement with our value derived from the W49\,N
sightline (in their study, the uncertain microwave-derived CH column densities were augmented by optical
measurements). Even more remarkable is their significant correlation between the microwave CH and OH
emission-line areas; the data from the sightline to $\zeta$~Oph seem to suggest a bimodal distribution
of CH at low OH column densities. From a compilation of four sightlines and two dark clouds they conclude
that when the main gas-phase carrier of carbon is CO rather than C$^+$, the CH abundance declines
markedly, in contrast to that of OH. This may be another indication of a bimodal or dual-slope relationship
between $N(\mathrm{CH})$ and $N(\mathrm{H_2})$. 

We conclude this Section with a remarkable detail observed on the sightlines to W3(OH) and W3\,IRS\,5.
Although the illuminating hot cores are separated by only $16\farcm 6$ (9.7~pc), the column density
profiles are quite different, even in the local arm where only the sightline to W3(OH) sees a narrow,
$\sim$1~\kms -wide feature (unlike the sightline to W49\,N, there is no risk of confusion with other
spiral arms close to zero velocity). Assuming a length of at most
500~pc  for the path through the local arm \citep{Lallement2014}, this converts to an upper limit of $2.4$~pc for the diffuse cloud size.
A similar upper limit was estimated by \citet{Winkel2017} for the sightline to SgrB2~M/N.
Interestingly, \citet{Liszt2016} determine a clumping scale of 5.5~pc for the gas on the sightline
to W\,31C. It seems fair to say that these estimates refer to the largest scales in diffuse gas
entities, with presumably considerable sub-structure.
\begin{figure}[h!]
\centering
\resizebox{\columnwidth}{!}{\includegraphics{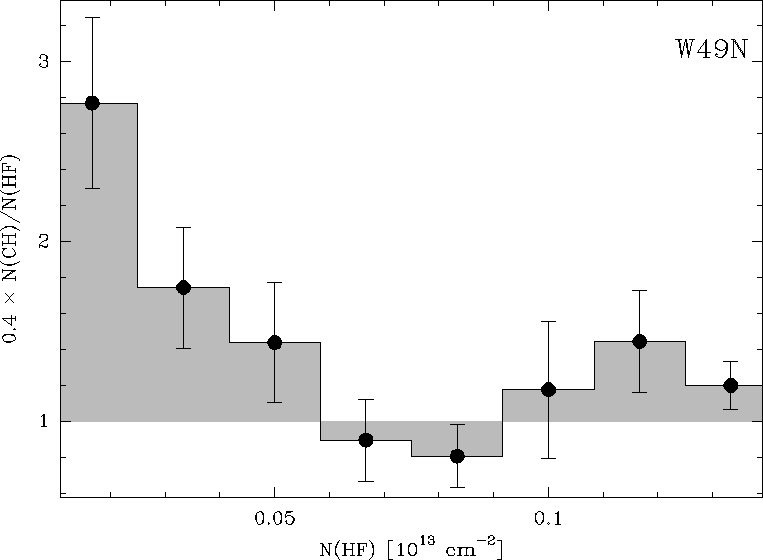}}
\resizebox{\columnwidth}{!}{\includegraphics{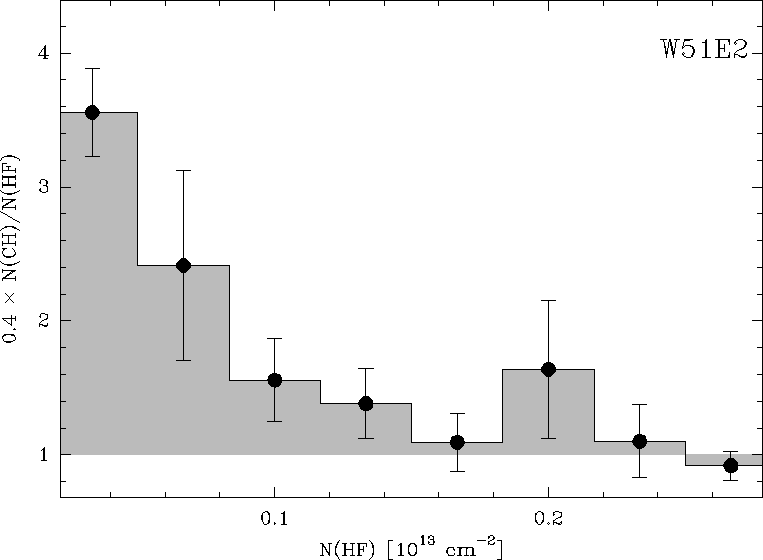}}
\caption{Over-abundance of CH, defined as $0.4 N(\mathrm{CH})/N(\mathrm{HF})$ vs. $N(\mathrm{HF})$.
         Column densities are derived from 1~\kms -wide velocity intervals in the spectral distribution
         of $N_{\rm v}$ (cf. Fig.~\ref{Fig2}). Only column densities with at least a $3\sigma$
         detection are used. Error-bars are deduced from the variances of the abundances
         within each histogram bin.
         \label{Fig6}
        }
\end{figure}
\begin{figure}[h!]
\centering
\resizebox{\columnwidth}{!}{\includegraphics{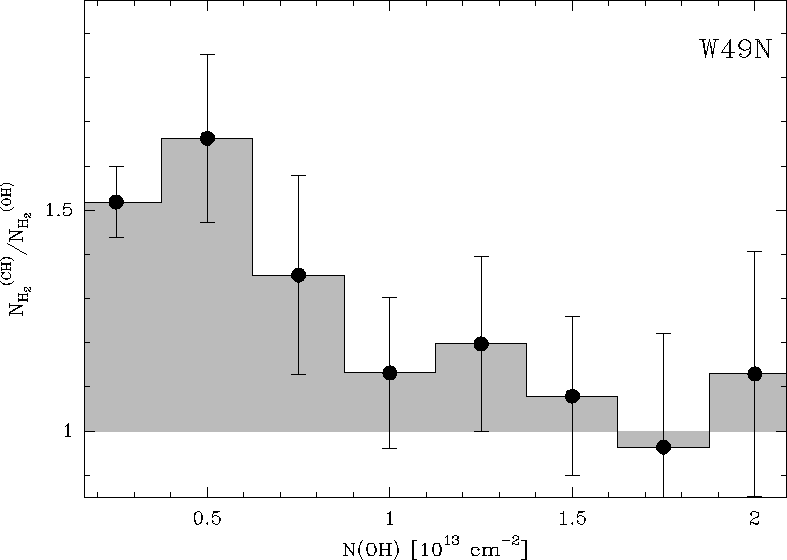}}
\caption{As in Fig.~\ref{Fig6}, but using OH as H$_2$ proxy instead of HF. The ordinate shows the
         ratio between H$_2$ column densities derived from CH and those derived from OH (assuming an
         $\mathrm{OH}/\mathrm{H_2}$ abundance of $10^{-7}$).
         \label{Fig7}
        }
\end{figure}
\begin{table}
\caption{Best-fit parameters of the model for CH emission/absorption towards W3\,IRS\,5.}
\centering
\begin{tabular}{l c c c c}
\hline\hline
Layer          & $\upsilon$            & $\Delta\upsilon$ & $N(\mathrm{CH})$                & $T_{\rm ex}$  \\
               & \multicolumn{2}{c}{[\kms]}   & $[10^{14}\,\,\mathrm{cm}^{-2}]$ & [K]           \\
\hline
 & & & & \\
Core emission (HC)     & $-37.5$ & 2.6 & 80.6  &  93.4 \\
                       & $-34.9$ & 4.3 & 62.6  & 133.9 \\
Common envelope (CE)   & $-39.7$ & 5.4 &  2.4  &  25.7 \\
                       & $-39.5$ & 1.6 &  0.9  &  23.5 \\
                       & $-38.1$ & 9.6 &  4.8  &  64.1 \\
Foreground screen (FG) & $-40.6$ & 6.5 &  1.2  &   3.1 \\
                       & $-39.5$ & 2.0 &  0.8  &   3.1 \\
                       & $-21.0$ & 1.3 &  0.2  &   3.1 \\
                       & $ +0.5$ & 2.7 &  0.1  &   3.1 \\
\hline
\end{tabular}
\tablefoot{Columns 2 and 3: Center velocity and {\sc FWHM} of Gauss-profile velocity components, respectively.
Column 4: $dN/d\upsilon$ integrated across velocity profile of component.
Components: HC -- Emission components, radiatively pumped by hot core dust (beam-filling factor 0.003). CE -- Three-component,
moderately excited common envelope around emission components (beam-filling factor 0.15). FG -- Cold foreground
screen, seen in absorption in either line.}
\label{table:3}
\end{table}
\subsection{CH as molecular gas tracer in environments forming high-mass stars}\label{sec:sfr}
The following Section demonstrates the usefulness of the CH ground state transitions in the 
high-density environments associated with sites of high-mass star formation.
Because of the saturated absorption in HF and OH towards the background sources, only the
optically thin line wings of these tracers are usable as tools for measuring the H$_2$ column density in the
star-forming environment, provided that they do not blend with unrelated foreground gas. As
already mentioned, the far-infrared lines of CH do not suffer from this drawback, and the combined
analysis of the $\lambda 149\,\,\mu$m and $\lambda 560\,\,\mu$m lines allows one to further constrain
the physical conditions characterizing the emission region. For OH, such an
approach was demonstrated by \citet{Csengeri2012}. In the following we try to extract a maximum of quantities
directly from the sub-millimeter/far-infrared spectroscopy of CH, with an indispensable minimum of assumptions.
For the far-infrared transition, only one lambda doublet component is available. Because the
spectroscopic properties of both components are not very different (Table~\ref{table:1}), this does not result
in a major lack of information; fitting both components merely represents an additional consistency check for the derived
column density of CH. Among the targets presented above, we select W3~IRS~5. Its location at a distance of
2.0~kpc in the second quadrant of the Galaxy in the Perseus arm avoids the confusion
arising in the $\lambda 560\,\mu$m line owing to its large hyperfine splitting and leading to a
blend of emission and absorption components originating from different locations on the sightline.

W3~IRS~5 is a bright infrared source discovered by \citet{Wynn-Williams1972} and
harbors a cluster of hyper-compact HII regions \citep[e.g.,][]{Wilson2003,vanderTak2005} powered by five proto-OB
stars \citep{Megeath2005}, which lend the object the designation of a ``Trapezium in its making'' by the
latter authors. The dust continuum emission of its components SMM1 to SMM5 was analyzed by \citet{Wang2013}; they
derive column densities of $N(\mathrm{H}_2) = 1.0 - 4.4 \times 10^{23}$~cm$^{-2}$, assuming a dust temperature of 150~K
and source diameters of $\sim$1$''$ (confirmed by a visiblity analysis of their {\it Submillimeter Array} (SMA)
data, see also \citealp{Wilson2003}).

Our model consists of an ensemble of hot cores (labeled {\it hc} in the following), corresponding to
SMM1 to SMM5, a common envelope ({\it ce}) associated with the entire W3\,IRS\,5 complex, and a foreground
screen ({\it fg}) representing the Perseus spiral arm and gas in the solar neighborhood.
The objective function of the model is given by 
\begin{equation}
\chi^2 = \left\langle w_{\lambda 560} \left (T^\mathrm{(mod)}_\mathrm{\lambda 560}-T^\mathrm{(obs)}_\mathrm{\lambda 560} \right )^2
                     +w_{\lambda 149} \left (T^\mathrm{(mod)}_{\lambda 149}-T^\mathrm{(obs)}_{\lambda 149} \right )^2 \right\rangle
,\end{equation} 
with (omitting the indices $\lambda 149$, $\lambda 560$)
\begin{equation}
T^\mathrm{(mod)} = \left\{\Phi_\mathrm{hc}\left(\Sigma_{\rm c}+\Sigma_\mathrm{hc}(1-\mathrm{e}^{-\tau_\mathrm{hc}})\right) \mathrm{e}^{-\tau_\mathrm{ce}}+\Phi_\mathrm{ce}\Sigma_\mathrm{ce}(1-\mathrm{e}^{-\tau_\mathrm{ce}}) \right\} \mathrm{e}^{-\tau_\mathrm{fg}}
\label{eq:model}
,\end{equation}
where $T^\mathrm{(mod)}$ and $T^\mathrm{(obs)}$ represent the modeled and observed brightness temperature of either line,  respectively.
The average is to be taken across the spectral bandpass of the full line profiles; $w_{\lambda 560}$ and $w_{\lambda 149}$ are
weight factors accounting for the individual signal-to-noise ratio in each transition. $\Phi_\mathrm{hc}$ and $\Phi_\mathrm{ce}$ are the
beam-filling factors of the hot core and common envelope, respectively (assuming that the hot dust continuum and the hot-core
emission fill the same beam fractions). The source functions $\Sigma_\mathrm{c}$ for the continuum
emission, and $\Sigma_\mathrm{hc}$ and $\Sigma_\mathrm{ce}$ for the hot core and common envelope line emission, respectively, are
expressed as Rayleigh-Jeans temperatures. We note that in Eq.~\ref{eq:model} the source functions are expressed
as unique, {\it equivalent} excitation temperatures, related to the actual source function profile along the sightline,
$\Sigma(z)$, by
\begin{equation}
T_\mathrm{ex, eqv} = \frac{\int_0^\mathrm{L} \Sigma(z)\kappa(z) \mathrm{e}^{-\int_\mathrm{z}^\mathrm{L}\kappa(z')dz'}dz}{1-\mathrm{e}^{-\int_0^\mathrm{L}\kappa(z) dz}}
,\end{equation}
where $L$ is the length of the sightline through the source.
We emphasize that a unique emissivity is not required for Eq.~\ref{eq:model} to be valid, that is, the opacity
coefficient $\kappa$ of a given component does not need to be constant. While the width and center velocities of the
Gaussian decomposition of the $dN/d\upsilon$ profile are well constrained by the $\lambda 149\,\mu$m absorption
spectrum, the degeneracy between the filling factors and excitation temperatures can only be lifted by the
dependence of the optical depths on the excitation. As shown in Fig.~\ref{fig:A2}, the $\lambda 560\,\mu$m
line is more sensitive to the excitation than the line at $\lambda 149\,\mu$m.
For a least-square fit to both CH lines, we apply the Metropolis algorithm (\citealp{Metropolis1953}, see
also \citealp{Press1992} for an extension to continuous minimization). A straightforward implementation
of the algorithm yields a satisfactory fit, although the convergence is slow. When the convergence
became quasi-linear, it was possible to accelerate it by minimization of residua \citep{Auer1987}.
\begin{figure}[h!]
\centering
\resizebox{\columnwidth}{!}{\includegraphics{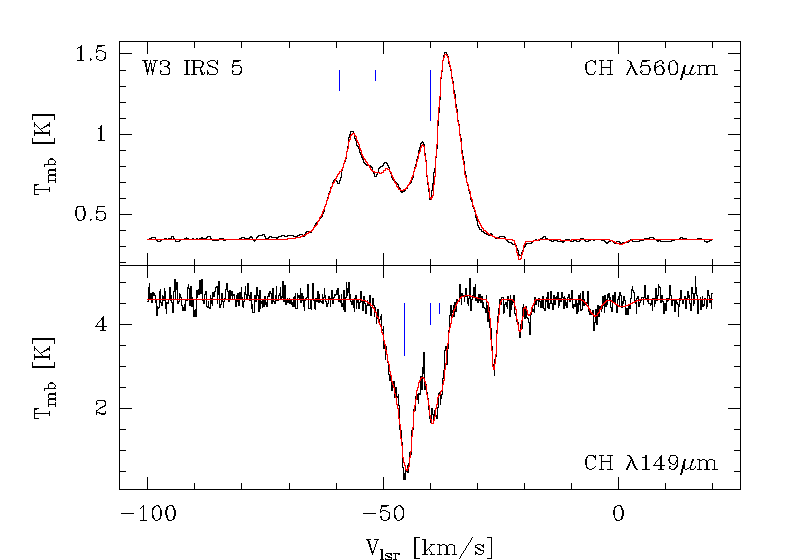}}
\caption{CH emission and absorption towards W3\,IRS\,5. Top: $\lambda 560\,\,\mu$m line (HIFI/PRISMAS).
         Bottom: $\lambda 149\,\,\mu$m (upGREAT). The red lines indicate the common fit to both spectra
         (in the $\chi^2$ sense). The hyperfine splitting is indicated by blue markers, scaling with
         the corresponding weight for a complete ground-state level population. The strongest hyperfine
         component of the $\lambda 560\,\,\mu$m line is centered close to the systemic velocity. For details
         see text.
         \label{Fig8}
        }
\end{figure}
The result is shown in Fig.~\ref{Fig8}, and the best-fit parameters are listed in Table~\ref{table:3}.
The normalized $\chi^2$ of our fit falls below 1\%, a result which could only be achieved by
assigning three components to the
warm, common envelope (likewise, various $\mathrm{H_2O}$ lines observed with HIFI were easier to fit
with a double component model, \citealp{Chavarria2010}). These components appear in absorption against
two highly excited emission components (presumably co-spatial with the hot core dust), otherwise in emission.
The prominent self-absorption at $-40$~\kms, in the strongest hyperfine-component of the $\lambda 560\,\mu$m
line, coincides with the radial velocity of the H$_2$O masers at $-40.7 \pm 1.3$~\kms
\citep[][, their group A outflow]{Imai2000}. Rather than treating the beam-filling factors in
Eq.~\ref{eq:model} as additional free parameters in
a high-dimensional parameter space, we fix them by means of ancillary continuum data: Identifying the
aforementioned compact, $\sim$1$''$ -wide dust sources in the SMA continuum image \citep{Wang2013} with
our hot emission components, the filling factor for the $39\farcs 5$ -wide HIFI beam (FWHM) amounts to
$\Phi_\mathrm{hc} = 0.003$. For the moderately warm common envelope, the SMA data
are less reliable for an estimate of $\Phi_\mathrm{ce}$, not least due to the lack of short spatial
frequencies in the continuum visibilities. We follow \citet{Chavarria2010} and use a
$\lambda 800\,\,\mu$m single-dish continuum map \citep{Oldham1994}, allowing us to estimate a
beam-deconvolved size of $\simeq 15''$, which converts to a beam filling of $\Phi_\mathrm{ce} = 0.15$.
Finally, two cold foreground components were needed for a satisfactory fit. Given their
velocities, they are probably related to the outer layers of the common envelope. The narrower
component (at $-39.5$~\kms) coincides with a common envelope component of comparable width.
This suggests that both components belong to the same structure, characterized by a negative excitation
gradient. Replacing these two components by a single one increases the normalized $\chi^2$ by 6\% and yields an
unsatisfactory fit in the $(-50,-40)$~\kms velocity interval, regardless of whether this component is
located in the common envelope or the cold foreground gas. As a matter of fact, we note that around 50~\kms
the fit to the $\lambda 560\,\,\mu$m line is still imperfect, even with the nine components ultimately retained.
Two further components at $\upsilon_\mathrm{lsr} = 0.5$ and $-21$~\kms are unrelated to W3\,IRS\,5 and can be
identified with local gas and with the near-side boundary of the Perseus spiral arm,
respectively. Thus, rather than freely fitting these two components, we kept their parameters fixed, adopting
the values from the analysis described in Section \ref{sec:diffuse}. 
\begin{figure}[h!]
\centering
\resizebox{\columnwidth}{!}{\includegraphics{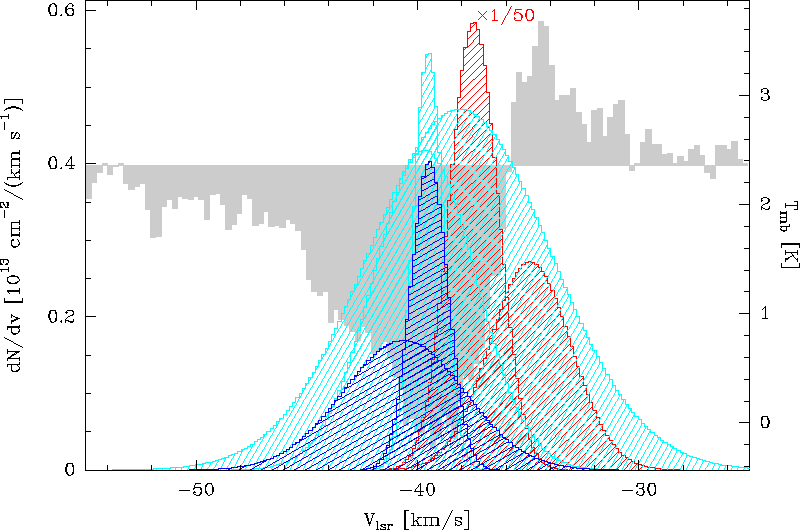}}
\caption{Breakup of CH components observed around W3\,IRS5. Red: $dN/d\upsilon$ for the two
         hot core components
         (scaled down by a factor of 50). Cyan: Same for the three common-envelope
         components. Blue: Cold foreground components. The gray histogram shows the P~Cygni profile
         of the HF ground-state transition ($J=1\rightarrow0$, archival HIFI data, PRISMAS key science
         project). For details see text and Table~\ref{table:3}. \label{Fig9}
        }
\end{figure}

Evidently, only a dedicated non-LTE simulation can explain the fitted excitation temperatures as a function of the physical
conditions encountered in the hot core and common envelope components. While this is beyond the scope of this paper,
we conclude this discussion with a comparison with the P Cygni-type profile seen in the HF $J=1\rightarrow 0$ line and
other light hydrides \citep{Benz2010}.
Figure~\ref{Fig9} shows the column density profiles for the various components against the HF spectrum. While the dominant
contribution to the total column density originates from the two hot core components, the prominent, saturated absorption feature
seen in HF is due to the same low-excitation material in the common envelope and foreground that is also traced by the absorption
seen in the CH lines. Our component separation clearly demonstrates that the bulk of this absorption
occurs at velocities that are blue-shifted with respect to the emission components. The driving agent for 
the expansion of the common envelope is most likely a system of multiple outflows \citep{Rodon2008}.
Another hint that in such an environment CH remains a reliable tracer for molecular hydrogen is provided
by the comparison of the derived column densities with corresponding estimates using the
$\lambda 850\,\,\mu$m dust emission \citep{Wang2013}: The H$_2$ column densities in the two emission
components amount to 2.3 and $1.8\times 10^{23}\,\mathrm{cm^{-2}}$ (for
$N(\mathrm{CH})/N(\mathrm{H_2}) = 3.5\times 10^{-8}$), which agree within a factor 2 with those derived
from the emission from the two dominant dust-continuum components
\citep[sources SMM1 and SMM2 in][]{Wang2013}. However, this good agreement may be a chance
coincidence: We have no proof that the best-fit solution summarized in Table~\ref{table:3} is
unique. The same holds for the rough dust-mass estimates (which, instead of a full continuum radiative
transfer, assume an ad-hoc dust temperature of 150~K).
\section{Conclusions}
We close this work with a summary of its main results:
\begin{enumerate}
\item 
For diffuse and translucent interstellar clouds, we confirm a prominent, but not very tight correlation
between the column densities of CH and HF. 
While a ratio of $N(\mathrm{CH})/N(\mathrm{HF}) \simeq 0.4$ represents rather a boundary 
to the observed relationship, a bimodal distribution of CH (with respect to HF) was found below
$N(\mathrm{CH}) \lesssim 10^{13}\,\mathrm{cm^{-2}}$ (i.e., $N(\mathrm{H_2}) \simeq 3\times 10^{20}\,\mathrm{cm}^{-2}$
or $A_{\rm V} \lesssim 0.3$). We identify the lower branch of this distribution
with the action of the endothermic reaction pathway, forming CH$^+$ thanks to the molecular hydrogen
abstraction reaction C$^+$(H$_2$,H)CH$^+$. The subsequent, fast hydrogen abstraction reaction forms
$\mathrm{CH}_2^+$, which is a prerequisite to form CH via dissociative recombination:
$\mathrm{CH}^+(\mathrm{H}_2,\mathrm{H})\mathrm{CH}_2^+(\mathrm{e}^-,\mathrm{H})\mathrm{CH}$. Because the
alternative pathway (forming CH$_2^+$ by a slow radiative association of C$^+$ with H$_2$) is less
efficient, the former pathway efficiently boosts the CH abundance. Our finding that this over-abundance
of CH appears in the lower-density regime strongly suggests that the agent to overcome the activation barrier
of 4640~K is dissipation of turbulence. Using OH instead of HF as a H$_2$ proxy yields the same conclusion
(albeit at lower significance). By the same token we determine an OH/CH abundance ratio of $2.37 \pm 0.15$.
\item In the envelopes surrounding cores forming high-mass stars, the combined analysis of the $\lambda 149\,\mu$m 
and $\lambda 560\,\mu$m ground state transitions of CH allows to separate emission features from absorbing layers
in the second transition, and therefore to better constrain both the column density and excitation of CH in
such environments. Follow-up studies will include a better determination of the beam-filling factors for the
far-infrared/sub-millimeter observations, by means of radio interferometry of the 3~GHz transitions between the hyperfine components
in the ground state. While these lines, fraught with non-LTE effects such as population inversion, will be difficult to interpret
without reliable collision coefficients at hand, radiative transfer models for the far-infrared/sub-millimeter lines
would be the next step in the analysis, for which this work may provide suitable boundary conditions.
\end{enumerate}

\begin{acknowledgements}
Based in part on observations made with the NASA/DLR Stratospheric
Observatory for Infrared Astronomy. SOFIA Science Mission Operations are
conducted jointly by the Universities Space Research Association, Inc., under
NASA contract NAS2-97001, and the Deutsches SOFIA Institut under DLR contract
50 OK 0901. We gratefully acknowledge the support by the observatory staff.
An anonymous referee provided valuable comments which improved the paper.
\end{acknowledgements}

\bibliographystyle{aa} 
\bibliography{references} 

\begin{appendix}
\section{Error estimates for double-sideband receiving systems}\label{app:err}
The determination of column densities from absorption spectra is
straightforward. This holds also for the case of hyperfine-split
spectra where the total column densities can be derived either by least-square
fitting of a correspondingly weighted target function, or by direct
deconvolution \citep[][Jacob et al., in preparation]{Gerin2010}. Nevertheless,
a candid discussion of measurement errors is essential. This statement holds
in particular for a double-sideband receiving system, where the calibration
of the continuum level is fraught with uncertainties arising from, for example, the
atmospheric transmission or the signal-to-image band gain
ratio. The measurement equation reads
\begin{equation}
T = T_{\rm c}\exp{(-\tau)} + \Delta T_{\rm c}\,,
\end{equation}
where $\Delta T_{\rm c}$ is an unknown contribution from the image band. While the
variance of $T_{\rm c}$ can be guessed from the unnormalized $\chi^2$ of a
least-square fit, $\Delta T_{\rm c}$ is a priori unknown, which is why the
target function is assumed to read
\begin{equation}
T = \left( T_{\rm c}+\Delta T_{\rm c} \right ) \exp{(-(\tau+\Delta\tau))}\,,
\end{equation}
where $\Delta\tau$ is the error resulting from the wrong attribution of
$\Delta T_{\rm c}$ to the signal band. Introducing $r=\Delta T_{\rm c}/T_{\rm c}$, one finds from the
above equations
\begin{equation}
\Delta\tau = \log{(1+r)}-\log{(r+\exp{(-\tau)})}-\tau\,,
\end{equation}
which is converted into a column density error thanks to Eq.~\ref{eq:1}.
The Bayesian error estimates throughout this paper are based on Monte-Carlo
simulations assuming Gaussian distributions of $T_{\rm c}$ (with $\sigma^2 = \chi^2$) and
$\Delta T_{\rm c}$. From a closer inspection of the atmospheric transmission in the signal and
image band of the $\lambda 149\,\,\mu$m line, and assuming typical calibration errors, we use
a Gaussian distribution of $\Delta T_{\rm c}$ with $\sigma_{\rm r} = 0.05 T_{\rm c}$.
The assessment of the resulting asymmetric errors will be discussed in the next section.

Our assumption of a 5\% uncertainty (r.m.s.) in the continuum level is motivated
by aspects related to hardware performance, and to the origin and transmission of the intrinsic
continuum emission. As for the hardware, hot electron bolometer mixers
\citep[e.g., ][]{Puetz2012} are characterized by signal-to-image band gain ratios close to unity.
\citet{Kester2017} provide a comprehensive study, showing that for HIFI bands 6 and 7,
using the same technology as upGreat, the uncertainty in the gain ratio falls below 4\%. At this
accuracy, there is no evidence for a departure from the theoretically expected 1:1 gain ratio.
Even a significant departure would not necessarily entail a corresponding calibration error as long as the spectral index of the continuum emission remains sufficiently small. The rest frequency of the far-infrared ground-state
transition of CH falls below the power-law part and the maximum of the spectral energy distribution of
$\sim$150~K hot dust. For dust opacity indices of 1.0 (expected for a predominant population of large
dust grains) and 2.0, the relative difference in the continuum emission in the signal and image band
amounts to 0.5 and 0.7\%, respectively, for the applied sideband separations of 3.4 and 3.6 GHz (for
W49\,N/W51\,e2, and W3\,OH/W3\,IRS\,5, respectively). The coupling efficiencies of the upGreat pixels are
monochromatic across the frequency range framed by the sidebands. Under the observing conditions described
in Section \ref{sec:observations}, the relative difference in the atmospheric signal and image band
transmissions amounts to at most 1\%. Quadratically adding the quoted errors yields a 4.2\% uncertainty,
falling short of the therefore conservative, retained 5\% level. For the HF $J=1\rightarrow 0$ line, which
was observed in HIFI band 5, a similar error estimate applies, despite the different technology (SIS mixers,
see, however, \citealp{Kester2017}). The overall calibration uncertainty, assumed to amount to $\sim$10\%,
is irrelevant here because we need only the line-to-continuum ratio.

\section{Assessment of asymmetric errors}\label{app:asymerr}
Column densities directly derived from substantial optical depths, which in turn are determined 
from the line-to-continuum ratio in absorption spectra, are fraught with asymmetric errors.
The accuracy of the line-to-continuum ratio (hereafter referred to as $r$) is limited by radiometric,
Gaussian baseline noise, and, sometimes, the uncertainty of the continuum level. This 
leads to uncertainties in the fitted line-to-continuum ratio, of which the non-linear transformation
to column densities lead to asymmetric errors. The analysis applied in this work follows
\citet{Barlow2003}: In his model (1) the errors in the derived quantity
are estimated by straight lines leaving from the central value, with slopes of $\sigma^+$ and $\sigma^-$
when the normal distribution of $r$ is transformed to a unit Gaussian. With the given
opacity mean $\tau_\mu = \langle \tau \rangle$, its variance $V=\langle \tau^2\rangle - \tau_\mu^2$ and the 
unnormalized skewness, $\gamma = \langle\tau^3\rangle-3\tau_\mu\langle \tau^2\rangle+2\tau_\mu^3$, at hand,
this leads to a system of three coupled, nonlinear equations for the median $\tau_{\rm m}$ and the
asymmetric errors $\sigma_{\tau,+}$ and $\sigma_{\tau,-}$, namely,
\begin{eqnarray}
\tau_{\rm m} & = & \tau_\mu-\frac{1}{\sqrt{2\pi}} \left (\sigma_{\tau,+}-\sigma_{\tau,-} \right )\,,\nonumber \\
f_1 & = & \frac{1}{4}\left ((\sigma_{\tau,+}+\sigma_{\tau,-})^2+(\sigma_{\tau,+}-\sigma_{\tau,-})^2(1-\frac{2}{\pi})\right )-V\,, \nonumber \\
f_2 & = & 2(\sigma_{\tau,+}^3-\sigma_{\tau,-}^3)-\frac{3}{2}(\sigma_{\tau,+}-\sigma_{\tau,-})(\sigma_{\tau,+}^2+\sigma_{\tau,-}^2) \nonumber \\
    &   & +\frac{1}{\pi}(\sigma_{\tau,+}-\sigma_{\tau,-})^3-\gamma\sqrt{2\pi}\,. \nonumber \\ 
\label{eq:asymmErrors1}
\end{eqnarray}
Model (2) of \citet{Barlow2003} considers a parabolic fit through the three
points (i.e., the central value and $\pm \sigma$), rather than the piecewise
defined straight lines of model (1). The above system of equations for
$\sigma_{\tau,\pm}$ and $\tau_{\rm m}$ then reads
\begin{eqnarray}
\tau_{\rm m} & = &\tau_\mu-\frac{1}{2} \left (\sigma_{\tau,+}-\sigma_{\tau,-} \right )\,,\nonumber \\
f_1 & = &\frac{1}{2}(\sigma_{\tau,+}+\sigma_{\tau,-})^2+\frac{1}{2}(\sigma_{\tau,+}-\sigma_{\tau,-})^2-V\,, \nonumber \\
f_2 & = &\frac{3}{4}(\sigma_{\tau,+}+\sigma_{\tau,-})(\sigma_{\tau,+}-\sigma_{\tau,-})+(\sigma_{\tau,+}-\sigma_{\tau,-})^3 -\gamma\,. \nonumber \\
\label{eq:asymmErrors2}
\end{eqnarray}
The solution of Equations~\ref{eq:asymmErrors1} or \ref{eq:asymmErrors2} is
determined (iteratively or graphically) by the crossings of the
$f_1(\sigma_{\tau,+},\sigma_{\tau,-})=0$ and the $f_2(\sigma_{\tau,+},\sigma_{\tau,-})= 0$ contours.
The resulting $(\sigma_{\tau,+},\sigma_{\tau,-})$ pair is subsequently used
to determine $\tau_{\rm m}$ from the first equation of the set \ref{eq:asymmErrors1} or
\ref{eq:asymmErrors2} (for model (1) and model (2), respectively).

\begin{figure}[h!]
\centering
\resizebox{\columnwidth}{!}{\includegraphics{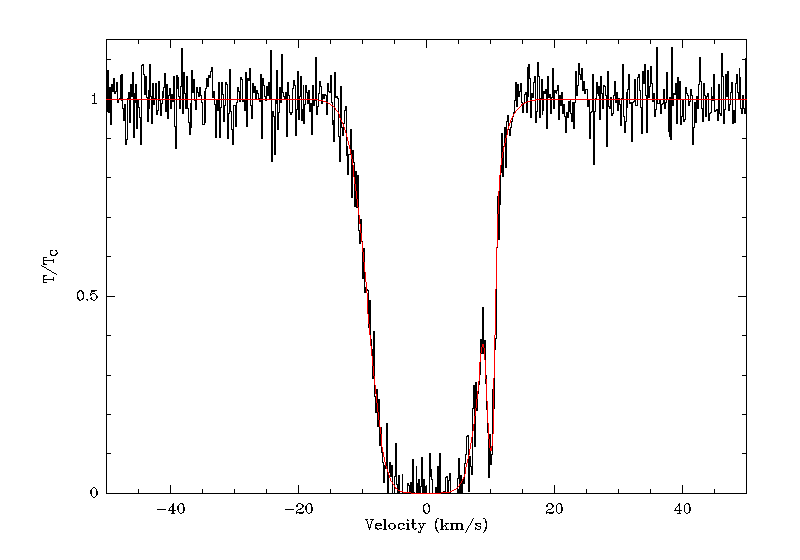}}
\resizebox{\columnwidth}{!}{\includegraphics{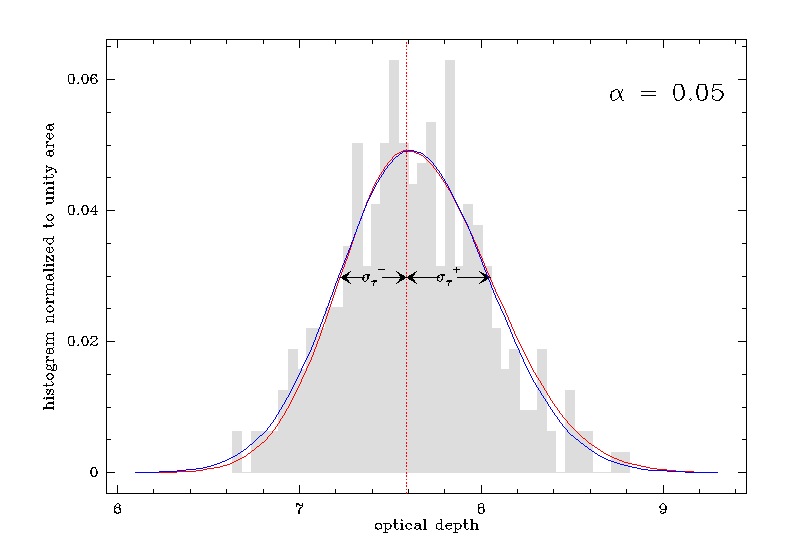}}
\resizebox{\columnwidth}{!}{\includegraphics{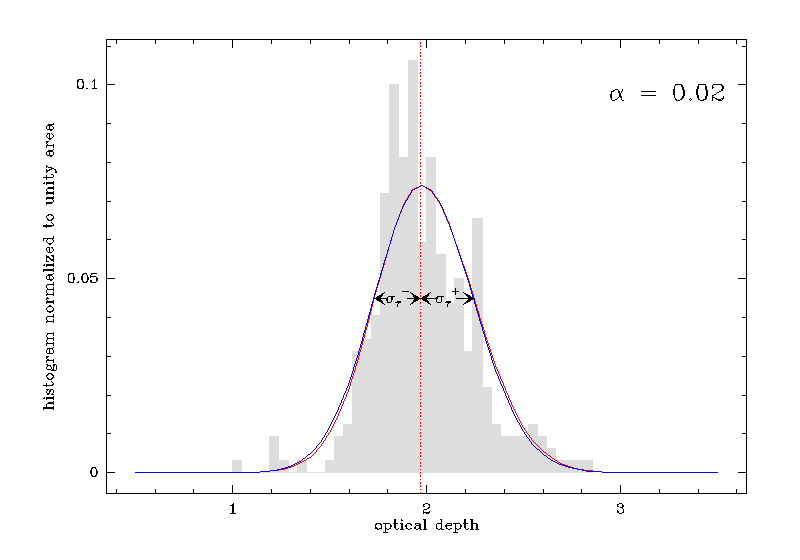}}
\caption{Demonstration of Bayesian error estimation according to
         \citet{Barlow2003}. Top: Test profile. The signal-to-noise ratio
         of the continuum is 20. Center and bottom: Histograms of the
         optical depth distribution, for the saturated absorption (center)
         and the narrower, unsaturated component (bottom). The dimidated
         Gaussians from models (1) and (2) are shown in red and blue,
         respectively, along with the asymmetric errors (arrows) and the
         asymmetry parameters $\alpha$. \label{fig:A1}}
\end{figure}

A demonstration of this approach is shown in Fig.~\ref{fig:A1}, for a
two-component opacity profile (a saturated absorption of $\tau_0 = 8$, and 
an unsaturated yet substantial one of $\tau_0 = 2$, located in the wing of the 
stronger absorption).  We note that in both cases the median value determined with
the above models is within $1\sigma$ limits from the original values, with the
stronger deviation for the saturated case. We also note that the line center
opacity in the saturated component could be determined thanks to the curvature
of the line profile in its unsaturated line wing opposite to the wing that
contains the unsaturated component. As expected, the asymmetry parameter 
$\alpha = (\sigma_{\tau,+}-\sigma_{\tau,-})/2$ is larger for the saturated
absorption component. One can also see that the difference between both models is almost
unnoticeable and irrelevant. Without loss of generality, the error
analysis applied in this paper relies on model (1).
\section{Comparison of column densities derived from the $\lambda 560$~\micron
and $\lambda 149$~\micron transitions}\label{app:chch}
\begin{figure}[h!]
\centering
\resizebox{\columnwidth}{!}{\includegraphics{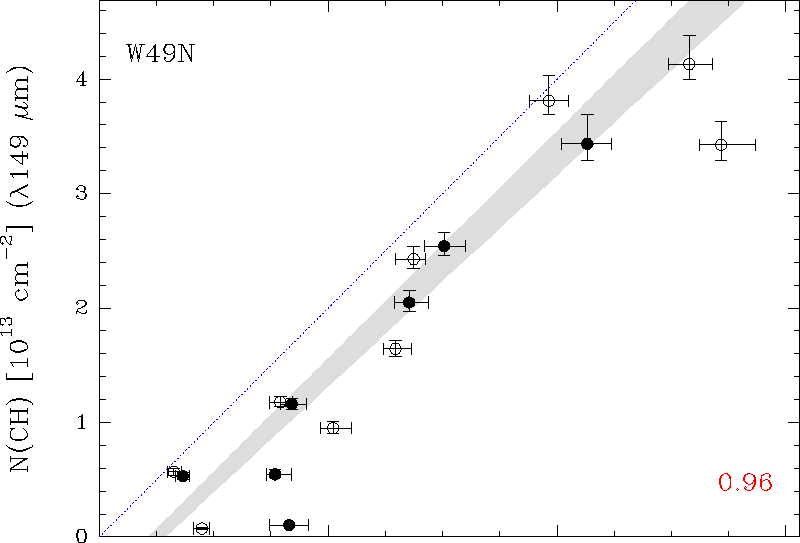}}
\resizebox{\columnwidth}{!}{\includegraphics{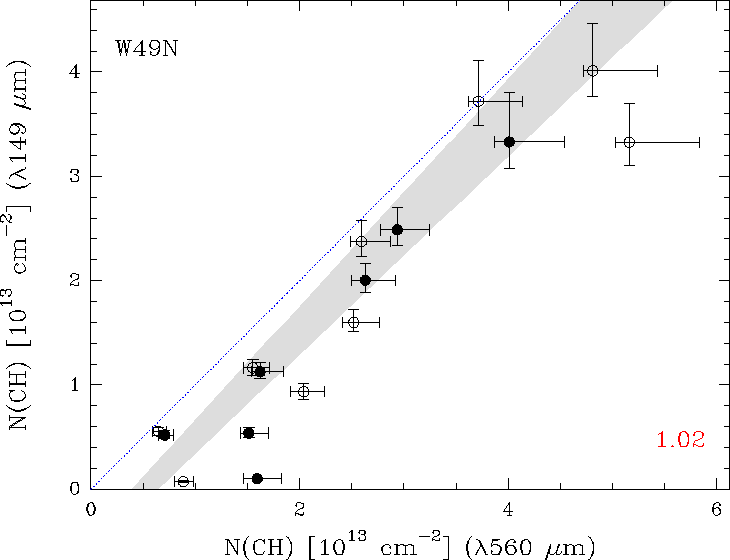}}
\resizebox{\columnwidth}{!}{\includegraphics{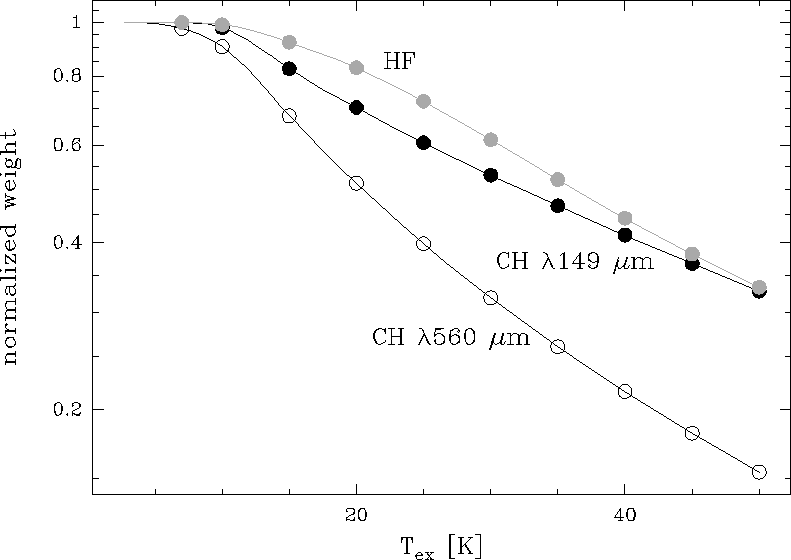}}
\caption{Comparison of column densities derived from the $\lambda 149$~\micron
         and $\lambda 560$~\micron lines, assuming a 5\% error in the continuum level
         (top) and a 10\% error (center). The gray-shaded area covers all data points within
         the 1$\sigma$ limits of the weighted linear regression (derived slopes in the bottom
         right corners). The column densities refer to 3~km/s wide velocity intervals in the
         spectral distribution of $N_{\rm v}$ (cf. Fig.~\ref{Fig2}). Only velocities above 
         30~km/s were considered. The dotted blue line indicates the bisector for identical
         column densities derived from both lines. Bottom: Normalized statistical weights
         to convert the total column densities to opacities. \label{fig:A2}
        }
\end{figure}

A comparison of CH column densities derived from the $\lambda 149$~\micron and
the $\lambda 560$~\micron lines may be instructive here. The weighted linear regression
is shown in Fig.~\ref{fig:A2}. Assuming a 5\% error in the continuum level,
the column densities derived from the $\lambda 149$~\micron line are 
systematically $(-0.40\pm 0.05)\times 10^{13}$~cm$^{-2}$ below those derived from the
$\lambda 560$~\micron line, with a slope of $(0.93 \pm 0.03) \times 10^{13}$~cm$^{-2}$
($\lambda 149$~\micron vs. $\lambda 560$~\micron). It seems fair to say that the explanation
for this discrepancy is either one of the following:
(1) As stated in the introduction, the determination of column densities
derived from the $\lambda 560$~\micron spectra needs to make an ad-hoc
assumption about the emission profile. Due to the mix of emission and absorption
around the hot core, the emission component is difficult to define, but
subsequently needed to disentangle the sightline absorption from the emission
component (which folds into the absorption by the spiral arms owing to the
relatively large hyperfine splitting). (2) Any inconsistency in the calibration
of the continuum level (e.g., of the contribution from the image band), with
respect to that of the spectral line, will lead to a corresponding error in the 
opacity determination (cf. the discussion in Appendix~\ref{app:err}).  The common
assumption that all CH molecules in the diffuse gas are in the ground state is
unlikely to be invalid here, as shown in Table~\ref{table:A1}. Without a deeper
analysis, it will therefore be difficult to identify the origin of the
discrepancy, and it seems both suggestions made above are at work.
\section{Non-LTE effects in the hyper-fine split CH ground state}\label{app:nlte}
Here we provide a short demonstration of NLTE effects in the ground state level populations
of CH under conditions typical for diffuse clouds. Table~\ref{table:A1} lists the departure
coefficients for two models, one corresponding to a diffuse cloud with low molecular content, 
the other one to a translucent cloud with a substantial fraction of molecules; the
calculations have been performed with the {\sc molpop-cep} code \citep{Elitzur2006},
assuming a simple slab model. All rotational CH levels up to $E_{\rm u} = 1120$~K were
considered. This implies radiative pumping, externally by the interstellar
radiation field (ISRF), as given by \citet{Mathis1983} and \citet{Draine1978}, and by the cosmic
microwave background, and internally by dust heated by the ISRF (resulting dust temperatures
are from \citealp{Voshchinnikov1999}). The quantities defining the clouds are from
\citet{Snow2006}. The escape probability is from \citet{Capriotti1965}. Alternatively,
the exact CEP solution was used, the relative differences in the level populations amount to
at most $1-3$~\%. For both techniques we assume a static cloud with a turbulent Doppler
width of 2~\kms (3.3~\kms FWHM). This choice is at the lower end of the linewidth-size
relation for diffuse gas on spatial scales from 10 to 100~pc \citep{Falgarone2009}
and close to the median width resulting from
fits to OH $^2\Pi_{3/2},\, J = 5/2 \rightarrow 3/2$ absorption line systems, observed on
nine sightlines through the first and fourth quadrant of the Galaxy \citep{Wiesemeyer2016}.
Collision rates are from \citet[][scaled from collisions with He to those
with H and H$_2$]{Marinakis2015}. The departure from LTE is model-dependent and varies within 
an order of magnitude.
\begin{table}
\caption{Conditions in diffuse cloud models and coefficients for
departure from LTE of the population in the CH $N=1,\, J=1/2$ HFS.}
\label{table:A1} \centering
\begin{tabular}{l c c}
\hline\hline
                & Model 1           & Model 2        \\
\hline
                & diffuse molecular  & translucent    \\
$\chi$ [Habing] & 1.7               & 1.7             \\
$A_{\rm V}$     & 0.2               &    1           \\
$n_{\rm H}$ [cm$^{-3}$]     & 100               & 1000           \\
$f^{\rm n}_{\rm H_2}$ & 0.1           &  0.5           \\
$T_{\rm gas}$ [K]     & 100               &   15           \\
$T_{\rm dust}$[K]     &  16               &   12           \\
$\Delta\upsilon_\mathrm{FWHM}$ [km/s] & 3.3             &   3.3          \\
$N(\mathrm{CH})$ [$10^{13}$~cm$^{-2}$]     & 1.75 & 1.58 \\
\hline
\multicolumn{3}{c}{LTE departure coefficients}  \\
\hline
$F=0+$ & 3.702 & 0.5480 \\
$F=1+$ & 3.721 & 0.5492 \\
$F=0-$ & 6.875 & 2.178  \\
$F=1-$ & 6.948 & 2.199  \\
\hline
\multicolumn{3}{c}{fractional population (per sub-level) in ground state}  \\
\hline
$F=0+$ & 0.0870 & 0.0504  \\
$F=1+$ & 0.0875 & 0.0505  \\
$F=0-$ & 0.1613 & 0.1981  \\
$F=1-$ & 0.1630 & 0.2000  \\
\hline
\multicolumn{3}{c}{fractional population in excited states [ppm]}  \\
\hline
  & 100 & 200 \\
\hline
\end{tabular}
\tablefoot{
Departure coefficients are defined as the fractional level population with respect to
LTE. The FUV field is parametrized in units of the Habing field
\citep[Draine field = 1.7~Habing,][]{Draine1978}.}
\end{table}
\end{appendix}
\end{document}